\documentclass[twocolumn,nofootinbib,prd,floatfix,preprintnumbers,amsmath,amssymb,groupedaddress]{revtex4}
\usepackage{dsfont}
\usepackage{epsfig}
\usepackage{slashed}
\usepackage{bbold}
\usepackage{feynmp}
\usepackage{psfrag}
\usepackage{color}
\PassOptionsToPackage{caption=false}{subfig}
\usepackage{subfig}
\usepackage{multirow}
\usepackage{booktabs}

\newcommand{\beq}{\begin{equation}}
\newcommand{\eeq}{\end{equation}}
\newcommand{\bea}{\begin{eqnarray}}
\newcommand{\eea}{\end{eqnarray}}
\definecolor{MMAblue}{rgb}{0.2472, 0.24, 0.6}
\definecolor{MMAred}{rgb}{1, 0, 0}
\definecolor{MMAgreen}{rgb}{0, 0.3333333, 0}
\definecolor{MMAorange}{rgb}{1, 0.5, 0}

\begin{document}
\psfrag{cs}{$\frac{M^2}{\Phi}\frac{d\Phi}{d m_T^2}$}
\psfrag{cs2}{$\frac{M^2}{\Phi}\frac{d\Phi}{d m_{T2}^2}$}
\psfrag{cs22}{$\frac{M^2}{\Phi\Phi}\frac{d\Phi\Phi}{d m_T^2}$}
\psfrag{mT}{$\frac{m_T^2}{M^2}$}
\psfrag{mT2}{$\frac{m_{T2}^2}{M^2}$}
\setlength{\unitlength}{1mm}
\title{Counting dark matter particles in LHC events}

\author{Gian Francesco Giudice}
\affiliation{Theory Division, CERN, 1211 Geneva 23, Switzerland.}

\author{Ben Gripaios}
\affiliation{Theory Division, CERN, 1211 Geneva 23, Switzerland.}

\author{Rakhi Mahbubani}
\affiliation{Theory Division, CERN, 1211 Geneva 23, Switzerland.}
%\pacs{PACS}
%\keywords{keywords}
\preprint{CERN-PH-TH/2011-195}
\begin{abstract}
We suggest trying to count the number of invisible particles produced in missing energy events at the LHC, arguing that multiple production of such particles provides evidence that they constitute stable Dark Matter and that counting them could yield further insights into the nature of Dark Matter. We propose a method to count invisible particles, based on fitting the shapes of certain transverse- or invariant-mass distributions, discuss various effects that may affect the measurement, and simulate the use of the method to count neutrinos in Standard Model processes and Dark Matter candidates in new physics processes.
\end{abstract}
\maketitle
\tableofcontents
%%%%%%%%%%%%%%%%%%%%%%%%%%%%%%%%%%%%%%%%%%%%%%
\section{Introduction}
%%%%%%%%%%%%%%%%%%%%%%%%%%%%%%%%%%%%%%%%%%%%%%
What a fillip it would be if the Dark Matter that abounds in the heavens could be manufactured here on Earth, at the LHC.
A Dark Matter (DM) particle, being colour singlet and electrically neutral, would be invisible in the LHC detectors, but would manifest itself in the form of ``missing energy'', recoiling against visible matter. 

While production of DM would imply a missing energy signal, the converse is not obviously true. For example, an invisible particle produced at the LHC need only live long enough to escape the detectors, while the DM in the cosmos has been around for billions of years. 
The question would thus arise of how to establish the link between a new, invisible particle produced at the LHC and the DM in the cosmos. 

One way would be to measure the properties of the invisible particle (and any new companions) and thence to compute the relic density and compare with cosmological observations. Alas, this is unlikely to be feasible at the LHC: even if we neglect the difficulties associated with making precision measurements at hadron colliders, there remains a basic obstacle, in
that while the relic density is determined (presumably) by the weak interactions of new particles, production of those same particles at hadron colliders proceeds dominantly via the strong interaction.

Here we should like to make a more modest, but perhaps more feasible, proposal for strengthening the hypothesis that a new, invisible particle produced at the LHC really is DM. Our proposal is simply to count the number of invisible particles in missing energy events. To begin with, our system of counting will be loosely based on the ``one-two-many'' system of the Amazonian Pirah\~{a} tribe \cite{Gordon:2004}, but simplified to ``one-many''. That is, we propose to try to establish that invisible particles are being multiply produced in events. 
As we argue in the next Section, this constitutes evidence for the existence of a symmetry that stabilizes DM. 

If we can establish that new invisible particles are being multiply produced, then it makes sense to follow the lead of the Pirah\~{a} and to try to extend our system of counting beyond ``one-many''. For example, we might try to count modulo some integer. We shall see, for example, that even counting only modulo two may lead to further insights about the nature of DM particles (and antiparticles) and the symmetry that stabilizes them.

Having argued, in \S\S \ref{sec:mult} and \ref{sec:count}, that counting invisible particles (or at least establishing that they are multiply produced) is a worthwhile thing to do at the LHC, we sketch, beginning in \S\ref{sec:light}, a method by which one might hope to do it. In a nutshell, our strategy is to identify observables that depend strongly on the number of invisible particles present, but which are relatively insensitive to all the other variables over which, {\em a priori}, we have little control. 
We begin by considering, for simplicity, the limiting case in which the masses of the invisible particles are negligible on collider scales. In \S\ref{sec:single}, we discuss the case in which new coloured particles are singly produced at the LHC and subsequently decay to visible and invisible particles; in \S\ref{sec:pair}, we discuss the case of pair production. In \S\ref{sec:mass},
we discuss various complications that arise in the massive case.
 In \S\ref{sec:other}, we discuss a variety of other effects, including matrix elements, finite widths, backgrounds and upstream transverse momentum.
 In \S\ref{sec:examples}, we apply our method  (neglecting showering and detector effects) to three Standard Model test cases, namely decays of a $W$-boson to a charged lepton and a single neutrino, decays of the Higgs boson to a pair of charged leptons and a pair of neutrinos via intermediate $W$-bosons, and decays of a pair of top quarks in the di-leptonic channel, and to a supersymmetric decay chain. Section \ref{sec:disc} provides a summary of the main results, while a number of the more cumbersome formul{\ae} have been relegated to an Appendix.
%%%%%%%%%%%%%%%%%%%%%%%%%%%%%%%%%%%%%%%%%%%%%%%%%%%%%%%%
\section{Multiple Production and Stabilizing Symmetries\label{sec:mult}}
%%%%%%%%%%%%%%%%%%%%%%%%%%%%%%%%%%%%%%%%%%%%%%%%%%%%%%%%
Why is multiple production of invisible particles evidence for DM? The obvious way to make DM sufficiently long-lived is to stabilize it by means of a symmetry: If DM is the lightest particle transforming non-trivially under the action of the symmetry (henceforth ``non-singlet"), then it cannot decay.\footnote{There are obvious analogues among the known particles: the proton is stabilized, perhaps accidentally, by baryon number, the lightest neutrino by fermion number, and the electron by electric charge.}
Suppose (as seems likely) that the dark matter particle is heavier than the constituent quarks of the proton, such that the quarks are singlets. Then protons and pairs of protons are singlet states, as are the products of LHC $pp$ collisions. If we are lucky enough to produce a non-singlet DM particle, then such a final state must contain another non-singlet particle. In the simplest case, this would be a second DM particle (or its anti-particle), but it could also be a different particle. If it is different, and if it is both visible and stable on detector length scales, then we will not see multiple production of invisible particles, but we will see a spectacular charged track. In all other cases, we must end up with multiple invisible particles in the detector, in order to form a singlet final state.

Before going on, let us argue that the observation of multiply-produced invisible particles would not be a trivial result, in that the alternative hypothesis of a new, singly-produced, invisible particle can plausibly be entertained.\footnote{Neutrinos give a known example of invisible particles that can be singly produced, but such events are easily picked out, assuming conservation of lepton number, by the presence of net visible lepton number in the final state.} 
Indeed, one might worry that if one could draw a Feynman diagram leading to significant production of such a state in association with SM states at the LHC, then the crossed diagram with only the invisible particle in the initial state would imply a lifetime less than the time required to traverse an LHC detector. However, this is not necessarily the case.  

Consider for example a resonance with mass $M$, strongly produced at the LHC, which decays into an electrically-neutral and
colour-neutral particle of mass $m$. The crossed Feynman diagram, with virtual exchange of the resonance, can contribute to the decay of the
``invisible" particle through a process which, at worst, involves a three-body final state. Then the lifetime is $\tau \sim 256\pi^3M^4/m^5$.
Being produced with typical momentum of order $M$, the ``invisible" particle will travel an average distance $L\sim \tau M/m$ before decaying. A
predominantly invisible decay (say with $L>30$ metres) is obtained for $m<2~{\rm GeV}(M/{\rm TeV})^{\frac{5}{6}}$. This allows for a reasonable range of masses.
Note, in particular, that the particle can be sufficiently heavy to avoid constraints from stellar cooling. Moreover, the bound on $m$ gets weaker if
the production process occurs in association with heavy SM particles (such as $W$, $Z$, $t$, or $b$), since the decay of the invisible particle must
involve multi-body final states. 

%%%%%%%%%%%%%%%%%%%%%%%%%%%%%%%%%%%%%%%%%%%%%%%%%%%%%%%%
\section{Counting Dark Matter \label{sec:count}}
%%%%%%%%%%%%%%%%%%%%%%%%%%%%%%%%%%%%%%%%%%%%%%%%%%%%%%%%
If we can establish that invisible particles are being multiply produced, we may also gain further insights by counting them, perhaps modulo some integer. 
As an example, if we could count the invisible particles modulo two, then observation of an odd number of invisible particles would rule out the simplest stabilizing symmetry, {\em viz.} $Z_2$. Indeed, 
an odd number of charged particles cannot form a singlet of $Z_2$. The simplest remaining candidate for a stabilizing symmetry would then be $Z_3$. An alternative approach to distinguishing $Z_2$ from other symmetries was discussed in \cite{Walker:2009ei,Agashe:2010tu}.\footnote{For models with stabilizing symmetries other than $Z_2$, see \cite{Agashe:2004ci,Lisanti:2007ec,Hambye:2008bq,Hambye:2009fg,Batell:2010bp,Adulpravitchai:2011ei}.} Moreover, if we make the plausible assumption that only the DM particle is stable on detector length-scales, then observation of an odd number of invisible particles also tells us that either the DM particle cannot be its own antiparticle, namely a real boson or a Majorana fermion, or that the stabilizing symmetry cannot be Abelian.

Indeed, any Abelian group is a product of $U(1)$s and finite cyclic groups
and DM must be charged under at least one of these. If, on the one hand, DM were its own antiparticle and were charged under a $U(1)$, one could not write the necessary mass term for DM in the Lagrangian. If, on the other hand, DM were charged only under some $Z_N$, one could write the mass term but could not simultaneously form a singlet final state from an odd number of DM particles.

We note, moreover, that counting modulo two is unaffected by the presence of neutrinos in the final state, since these too can be counted modulo two, via the visible lepton number in the final state. If we try to be even more ambitious and count invisible particles modulo $N>2$, then we should need to solve the problem of counting neutrinos on an event-by-event basis. One could imagine doing so, by using our knowledge of neutrino dynamics, but we shall not go into the details here. 
%%%%%%%%%%%%%%%%%%%%%%%%%%%%%%%%%%%%%%%%%%%%%%%
\section{Light invisible particles\label{sec:light}}
%%%%%%%%%%%%%%%%%%%%%%%%%%%%%%%%%%%%%%%%%%%%%%%%%%%%%%%%
%%%%%%%%%%%%%%%%%%%%%%%%%%%%%%%%%%%%%%%%%%%%%%%
\subsection{Single production\label{sec:single}}
%%%%%%%%%%%%%%%%%%%%%%%%%%%%%%%%%%%%%%%%%%%%%%%%%%%%%%%%
Having argued that counting invisible particles (or at least establishing that they are multiply produced) is a worthwhile thing to do at the LHC, let us now try to convince the reader that it can be done, at least in principle. To do so, we recall that invisible particles can be discovered by their recoil against visible particles, manifesting themselves in the form of missing energy in an event. Now, the total collision energy is shared out between all the visible and invisible particles in an event in a random fashion;  as a result, the shape of the distribution of the missing energy in a sample of many events will depend on the number of invisible particles present. 

In fact, the distribution of almost any observable will depend on the number of invisible particles present. This is, of course, good news. The bad news is that almost any observable will also depend on many other things, making it hard for one to be sure that one is really counting the number of invisible particles and not just measuring some poorly-defined combination of many other things. This leads us onto the important issue of how we propose to count invisible particles in practice.

To do so, one would like to find an observable which depends strongly (and in a known way) on the number of invisible particles, but which is less sensitive to all the other variables over which, {\em a priori}, we have little control. Examples of such pernicious variables include:
the sizes of Standard Model backgrounds, other unknown parameters in the new physics Lagrangian (such as particle masses and couplings), the details of the event topology, the unknown boosts of produced new particles with respect to the laboratory frame, uncertainties in the parton distribution functions (pdfs) \&c. There are other important effects to which we have less hope of being insensitive. These include: the unknown particle spins and widths, the details of hard process matrix elements, the presence of radiation in the initial state, detector resolution, and so on. Nevertheless, we shall argue that the sensitivity to these effects can be mitigated, for three reasons. One is that we can, of course, resort to numerical simulations of physics and detectors to try to model such effects. The second is that, as we shall see, the effect on certain distributions of changing the number of invisible particles is a dramatic one, and is unlikely to be masked or faked by other effects. The third reason is that, since we are trying to measure a discrete integer, we can tolerate a large error and still have confidence in our result. 

Before proceeding to a more detailed discussion of these issues, let us begin by considering a simple example that illustrates the dramatic effect on distributions that may result from changing the number of invisible particles. We consider decays of a single parent particle of mass $M$ into some number of visible and invisible daughter particles.
For the time being, we assume that the masses of the invisible particles are negligible at the scales of interest.
We should like to define an observable which has a strong dependence on the number of invisible particles, which can be understood in a simple way. We now claim that (a version of) the transverse mass \cite{vanNeerven:1982mz} is such an observable. We note that, given only a single visible particle, our observable could be any function of the momentum of the visible particle and the missing transverse momentum; with more visible particles, many more observables become possible, including, for example, the invariant mass of some of the visible particles. As described above, the distributions of any of these observables will depend on the number of invisible particles. But the transverse mass (and invariant masses, which we consider in \S \ref{sec:invmass}) are special among these because not only do their distributions have a maximal endpoint, which is easy enough to pick out in data, but also the condition for an event to be at the endpoint can be expressed in a simple way, in terms of the invisible particles. For example, in the case of the transverse mass, we shall see that necessary and sufficient conditions to be at the maximum are that all invisibles be parallel and transversely directed in the rest frame of the decay.\footnote{To avoid later confusion, we stress that the conditions change if the invisible particles have non-negligible mass.} As we shall see, the transverse mass has the additional advantage that its behaviour near its maximum is independent of whether or not the decay involves intermediate particles that are on-shell. 

To see all this, let us group the daughter particles of the decay into a visible system of transverse momentum $p_T$ and invariant mass $m_V$ and an invisible system (of $n$ particles) of transverse momentum $\slashed{p}_T$ and invariant mass $m_I$. The transverse mass is na\"{\i}vely obtained by projecting the invariant mass 
of all daughters onto the observable transverse directions, namely
\begin{multline} \label{naive}
m_{T,\mathrm{naive}}^2 \equiv \\ m_V^2+ m_I^2 + 2\left(\sqrt{(\slashed{p}_T^2+ m_I^2)(p_T^2+m_V^2)} - \slashed{p}_T \cdot p_T\right).
\end{multline}
However, the invariant mass of the invisible system, $m_I$, is not observable and thus nor is $m_{T,\mathrm{naive}}$. However, we can simply replace the true value of $m_I$ by some fixed value in the definition.
We choose to replace it by its minimum value (which, since $m_{T,\mathrm{naive}}$ is a monotonically increasing function of $m_I$, preserves the property $m_{T} < M$). This minimum value is given by the sum of the masses of the invisible particles,
which we have taken, for now, to be negligible.
We thus arrive at our final definition of the transverse mass, namely\footnote{The transverse mass thus defined is, in fact, the natural variable to use in this case, in the sense that it captures all of the information that is available from kinematics alone \cite{Barr:2009jv}.}
\begin{gather} \label{mt}
m_{T}^2 \equiv m_V^2+ 2\left(\sqrt{\slashed{p}_T^2(p_T^2+m_V^2)} - \slashed{p}_T \cdot p_T\right).
\end{gather}
We now examine the conditions that events must satisfy, in order to be at the maximum endpoint of the $m_T$ distribution. Since the invariant mass of the parent may be expressed as
 \begin{gather}
M^2 = m_V^2+ m_I^2 + 2(E_V E_I - \slashed{p}_T \cdot p_T   - q_V q_I),
\end{gather}
where $E$ denotes the energy and $q$ the longitudinal momentum component,
we have that
 \begin{multline} \label{mtM}
m_{T}^2 = M^2 - m_I^2 -  2 e_V\left(\sqrt{\slashed{p}_T^2+ m_I^2}-\sqrt{\slashed{p}_T^2}\right) \\ -2(E_V E_I - e_V e_I - q_V q_I),
\end{multline}
where we defined the transverse energies via $e^2 \equiv p_T^2 + m^2$. 
Using the relation
 \begin{gather} \label{eqn:rel}
(E_V E_I)^2 - (e_V e_I + q_V q_I)^2 = (e_V q_I - q_V e_I)^2,
\end{gather}
it is clear that $E_V E_I \geq e_V e_I + q_V q_I$. But then the last three terms on the right hand side of (\ref{mtM}) are each negative semi-definite, such that
 \begin{gather} 
m_{T}^2 \leq M^2,
\end{gather}
with equality iff.\ 
 \begin{gather} 
e_V q_I - q_V e_I = 0 \; \;\; \mathrm{and}  \; \; \;m_I =0.
\end{gather}
The former condition is invariant under longitudinal Lorentz boosts and implies either that $q_V = q_I=0$ or that the configuration is a longitudinal boost thereof.
The latter condition is invariant under all Lorentz boosts and implies that all the invisible particles' momenta are parallel (not anti-parallel). Thus, the condition to be at the maximum of the $m_T$ distribution is that all invisibles be parallel, and that there exist a longitudinal boost to a frame in which they be purely transverse. (The generalization to the case of massive invisibles will be carried out in \S\ref{sec:mass}.)

\begin{figure}[t]
  \psfrag{1}{\color{MMAblue}\boldmath$n\!\!=\!\!1$}
  \psfrag{2}{\color{MMAred}\boldmath$n\!\!=\!\!2$}
  \psfrag{3}{\color{MMAgreen}\boldmath$n\!\!=\!\!3$}
  \psfrag{4}{\color{MMAorange}\boldmath$n\!\!=\!\!4$}
  \begin{center} 
    \includegraphics[width=0.99\linewidth]{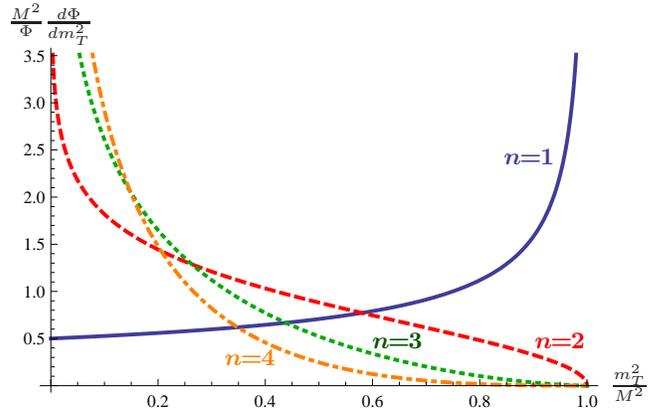}
    \end{center}
\caption{Normalized, phase space distribution of $m_T^2$ for $n \in \{1, \dots, 4\}$ for  the process $M \to P+nX$, {\it i.e.} the decay of a parent particle of mass $M$ into a visible particle $P$ and $n$ invisible particles $X$. \label{fig:mt}} 
\end{figure}

\begin{figure}[t]
  \psfrag{1}[br]{\color{MMAblue}\boldmath$n\!\!=\!\!1$}
  \psfrag{2}{\color{MMAred}\boldmath$n\!\!=\!\!2$}
  \psfrag{3}{\color{MMAgreen}\boldmath$n\!\!=\!\!3$}
  \psfrag{4}{\color{MMAorange}\boldmath$n\!\!=\!\!4$}
  \begin{center} 
    \includegraphics[width=0.99\linewidth]{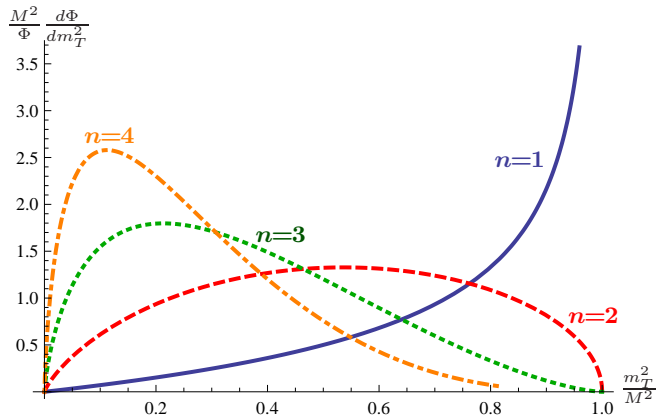}
    \end{center}
\caption{Normalized, phase-space distribution of $m_T^2$ for $n \in \{1, \dots, 4\}$ for  the process $M \to 2P+nX$, {\it i.e.} the decay of a parent particle of mass $M$ into two visible particles $P$ and $n$ invisible particles $X$. \label{fig:mt2}} 
\end{figure}

We thus see that the conditions to be at the maximum have a strong dependence on $n$ (since one must line up all $n$ invisibles), but are independent of (i) the number of visibles and their masses and (ii) longitudinal boosts of the parent with respect to the lab frame. We can guess the form of the $m_T$ distribution near its endpoint, for configurations in which the parent is produced without any transverse momentum and decays according to phase space considerations alone, in the following way. The dependence coming from the condition that the total momentum of the invisibles be purely transverse is independent of $n$ and may be evaluated at $n=1$; it comes from the Jacobian that arises in changing variables from the decay direction (which is uniformly distributed) to $m_T^2$ and gives a factor of $ \left(1- \frac{m_T^2}{M^2}\right)^{-\frac{1}{2}}$. The $n$ dependence comes from the distribution of the invariant mass of the invisible particles, which, as can be seen in (\ref{eqn:phs}) goes as $m_I^{2(n-2)}$. Near the endpoint (at small $m_I^2$), $m_T^2$ is linear in $m_I^2$ and hence we get a factor of $\left(1- \frac{m_T^2}{M^2}\right)^{n-2}$. In all, we find
\begin{gather} \label{univ}
\frac{d\Phi_{n+1}}{dm_T^2} \propto \left(1- \frac{m_T^2}{M^2}\right)^{n-\frac{3}{2}}.
\end{gather}
We reproduce the explicit calculation of the $m_T$ distribution for the cases of one and two massless visible particles in the Appendix; 
 the resulting expressions are given in (\ref{eqn:mtdist11}-\ref{eqn:mtn14}) and  (\ref{eqn:mtn241}-\ref{eqn:mtn24})
and distributions for the first few values of $n$ are illustrated in Figs.~\ref{fig:mt} and \ref{fig:mt2}.
%%%%%%%%%%%%%%%%%%%%%%%%%%%%%%%%%%%%%%%%%%%%%%
%%%%%%%%%%%%%%%%%%%%%%%%%%%%%%%%%%%%%%%%%%%%%%
We see that the dependence on the number of particles is, in some sense, the strongest possible: for $n=1$, the distribution diverges at the endpoint (in the absence of finite experimental resolution), whilst it vanishes for $n>1$. So we can hope that it is possible to discriminate single and multiple production of invisible particles, by looking at the endpoint behaviour. (As regards the secondary question of counting the number of invisibles for $n>1$, the fall-off with $n$ is so rapid that a lack of statistics near the endpoint would be a concern for $n$ greater than a few.)

To see whether such a hope could become a reality at the LHC, one must consider the various aforementioned complications. Happily, the result is robust with respect to a number of effects. We have already argued that, since the transverse mass depends on only transverse or Lorentz-invariant quantities, the phase-space distribution will be insensitive to pdfs (which give rise only to longitudinal boosts of the singly-produced parent) and their uncertainties. 
%%%%%%%%%%%%%%%%%%%%%%%%%%%%%%%%%%%%%%%%%%%%%%
%%%%%%%%%%%%%%%%%%%%%%%%%%%%%%%%%%%%%%%%%%%%%%
We have also argued that, at least in the limit of massless invisibles, the endpoint behaviour is independent of the number of visible particles and their masses.\footnote{Even if there were a dependence, this would be easy to compensate for, since the number and masses of visible particles are observable.}
We now show that, in the same limit, the power-law fall-off in $m_T$ is also unchanged if we let intermediate states be on-shell (leading to, for example, cascade decays). 
%%%%%%%%%%%%%%%%%%%%%%%%%%%%%%%%%%%%%%%%%%%%%%
%%%%%%%%%%%%%%%%%%%%%%%%%%%%%%%%%%%%%%%%%%%%%%
\subsubsection{Topology dependence \label{sec:singlecas}}
%%%%%%%%%%%%%%%%%%%%%%%%%%%%%%%%%%%%%%%%%%%%%%
It is easy enough to convince oneself that the endpoint behaviour of the $m_T$ distribution will be unaffected by the presence of on-shell intermediate states, with one {\em caveat}. Indeed, we have already seen that necessary and sufficient conditions to be at the endpoint are that the massless, invisible particles be parallel ($m_I=0$) and transverse, $q_I=0$, regardless of the topology. The problem then reduces to studying how easy it is to satisfy these two conditions, as one varies the topology. The condition that the particles be transverse cannot depend on the topology, since the phase space for any topology is always a Lorentz invariant. For the condition that the particles be parallel, 
an explicit calculation confirms that $\frac{d \Phi}{d m_I^2} \sim m_I^{2(n-2)}$ as $m_I^2 \rightarrow 0$ independently of the topology, with one exception. The exception
is that there exist topologies for which one cannot hope to satisfy $m_I=0$, namely those in which an on-shell intermediate decays exclusively into invisible particles. Then, the invisible particles from that decay cannot all be parallel, if energy and momentum are to be conserved.\footnote{On reflection, such a situation must result in a pathology, since even the most diligent experimentalist could not be privy to the knowledge that an invisible intermediate particle had decayed into other invisible particles. Clearly, what would happen in practice would be that the intermediate particle would be counted as a single (massive) invisible particle.} In all other cases, the endpoint behaviour of the $m_T$ distribution will be unaffected by the presence of on-shell intermediates.

\begin{figure}[t]
  \psfrag{0}{\color{MMAblue}\boldmath$z\!\!=\!\!\infty$}
  \psfrag{0.01}{\color{MMAred}\boldmath$z\!\!=\!\!0.01$}
  \psfrag{0.5 }[br]{\color{MMAgreen}\boldmath$z\!\!=\!\!0.5$}
  \psfrag{0.99}[br]{\color{MMAorange}\boldmath$z\!\!=\!\!0.99$}
  \begin{center} 
    \includegraphics[width=0.99\linewidth]{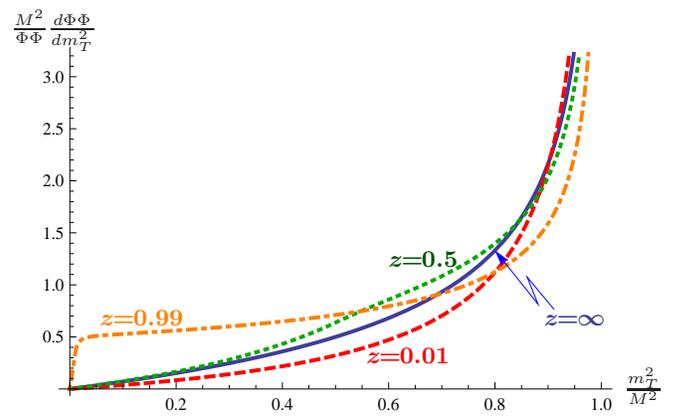}
    \end{center}
\caption{Normalized, phase-space distribution of $m_T^2$ for the
  cascade decay process $M \to P+Y \to 2P + X$ for various values of the
  intermediate mass ratio $z=m_Y^2/M^2$. \label{fig:mt_cascade}} 
\end{figure}

In (\ref{eqn:casc}), we give the explicit form of the distribution in
the simplest case of a decay involving two massless, visible particles
and one massless, invisible particle, connected by a single
intermediate state of mass $m_Y$; one may easily show that the
endpoint behaviour is given by (\ref{univ}) for all $m_Y$. Distributions for various
intermediate mass ratios, $z=m_Y^2/M^2$, are shown in
Fig.~\ref{fig:mt_cascade}.
%%%%%%%%%%%%%%%%%%%%%%%%%%%%%%%%%%%%%%%%%%%%%%
%%%%%%%%%%%%%%%%%%%%%%%%%%%%%%%%%%%%%%%%%%%%%%
\subsubsection{Invariant mass observables \label{sec:invmass}}
%%%%%%%%%%%%%%%%%%%%%%%%%%%%%%%%%%%%%%%%%%%%%%
In cases where there are multiple visible particles present in the final state, one also has the option of using the invariant mass of some combination of the visible particles as an observable. Compared to the transverse mass, these have the advantage of being invariant under all boosts and we shall see that they too have endpoint behaviours that depend strongly on the number of invisible particles. Perhaps their most appealing property is that they do not require one to measure the missing energy in events, with the well-known difficulties that such a measurement entails.\footnote{Measurement of the missing energy might be required for triggering, or to establish the presence of a signal, however.}

\begin{figure}[t]
  \psfrag{n1}{\color{MMAblue}\boldmath$n\!\!=\!\!1$}
  \psfrag{n2}{\color{MMAred}\boldmath$n\!\!=\!\!2$}
  \psfrag{n3}{\color{MMAgreen}\boldmath$n\!\!=\!\!3$}
  \psfrag{n4}[br]{\color{MMAorange}\boldmath$n\!\!=\!\!4$}
 \psfrag{cs}{$\frac{M^2}{\Phi}\frac{d\Phi}{d m_V^2}$}
\psfrag{mT}{$\frac{m_V^2}{M^2}$}

  \begin{center} 
    \includegraphics[width=0.99\linewidth]{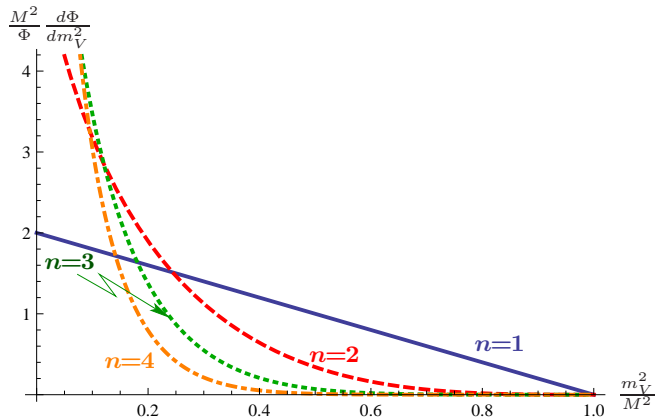}
    \end{center}
\caption{Normalized, phase-space distribution of $m_V^2$ for $n \in \{ 1, \dots, 4 \}$ for  the process $M \to2 P+nX$. \label{fig:mv}} 
\end{figure}

Let us proceed to discuss how the endpoint behaviour of such observables depends on the number of invisible particles. Consider the case in which a single particle of mass $M$ decays into $l$ massless, visible particles and $n$ massless, invisible particles. Now, it is simple to show that the invariant mass, $m_V$, of the $l$ visible particles reaches its maximum only when all $n$ invisible particles are produced with vanishing momentum in the rest-frame of the decaying particle.\footnote{Provided the decaying particle is boosted with respect to the lab frame, there will still be missing energy on which to trigger, if required.} Thus, if we add one extra invisible particle, the extra condition to remain at the maximum of $m_V$ is stronger than the extra condition to remain at the maximum of $m_T$; to wit, in the first case the extra particle must have vanishing momentum (in the rest frame of the decay), whilst in the second case it need only be parallel with the existing invisibles. As a result the coefficient of $n$ in the power-law fall-off near the endpoint is increased; in the simplest case $l=2$, an explicit calculation shows that\begin{gather} \label{mv}
\frac{d\Phi_{n+2}}{dm_V^2} \propto \left(1- \frac{m_V^2}{M^2}\right)^{2n-1}.
\end{gather}
We reproduce the full $m_V$ distribution in (\ref{eqn:minv2vni}); distributions for the first few values of $n$ are illustrated in Fig.~\ref{fig:mv}.
%%%%%%%%%%%%%%%%%%%%%%%%%%%%%%%%%%%%%%%%%%%%%%
%%%%%%%%%%%%%%%%%%%%%%%%%%%%%%%%%%%%%%%%%%%%%%

Invariant mass distributions have the apparent disadvantage that, unlike the $m_T$ distribution, their behaviour at their maximum does change when intermediate particles go on-shell and therefore use of them to count invisible particles requires one to first know (or assume) the event topology. This is hardly surprising, since the location of the endpoint itself changes as intermediate particles go on shell. To give an explicit example, in the simplest cascade decay discussed above with two visible particles and an on-shell intermediate of mass $m_Y$, the endpoint is at $m_V^2 = M^2 - m_Y^2$ and the distribution there is flat. But for a point-like three body decay, the maximum is at $m_V^2 = M^2$ and the distribution vanishes linearly there. 
%%%%%%%%%%%%%%%%%%%%%%%%%%%%%%%%%%%%%%%%%%%%%%
\subsection{Pair production \label{sec:pair}}
%%%%%%%%%%%%%%%%%%%%%%%%%%%%%%%%%%%%%%%%%%%%%%
We now switch our attention to decays of pair-produced particles. Pair production is relevant in theories, such as minimal supersymmetric models, in which the parent states are charged under the stabilizing symmetry and must themselves be multiply produced. It turns out that the $m_T$ distribution still has the same form of power-law fall-off for the case of pair-produced parents, assuming that the two parents are produced at rest, relative to one another, and without an overall transverse boost. These are, perhaps, not such good assumptions at the LHC where, even in the absence of initial state radiation, two parents may be produced boosted back-to-back in their rest frame.\footnote{Parents will necessarily be boosted if the production process involves non-zero angular momentum.} One way to avoid this problem would be to use, instead of $m_T$, the invariant mass of the visible decay products coming from one of the decaying parents. This observable is, of course, invariant under any boost of that parent, but the endpoint behaviour of its distribution will only yield information about the number of invisible particles produced in that parent's decay. Alternatively, one may use the distribution of the observable $m_{T2}$ \cite{Lester:1999tx,Barr:2003rg}. Again, this is the natural observable for pair production, in that it encodes all of the information that is available from kinematic considerations \cite{Serna:2008zk,Cheng:2008hk,Barr:2009jv}.\footnote{Here, we focus on pair production of parents of identical mass; generalizations of $m_{T2}$ \cite{Barr:2009jv} might be more appropriate in other cases.}
 More importantly as regards our discussion, it enjoys a large invariance with respect to boosts of the parent particles, reducing the sensitivity to the details of the production process. This invariance is exhibited most simply by noting that if we only consider events in which the invisible particles recoil against the visible particles in the decays, such that there is no initial state radiation or other source of transverse momentum ``upstream'' in the event, then the algorithmic definition of  $m_{T2}$ reduces to an algebraic expression \cite{Lester:2007fq,Cho:2007dh,Lester:2011nj}.
The expression depends on whether a configuration is 
``balanced'' or 
``unbalanced'' as defined in \cite{Lester:1999tx,Barr:2003rg} and is given by\footnote{As we did for $m_T$ ({\em cf.} (\ref{mt})), we have replaced the true, unknown value of the invariant mass of the invisible system in each decay with zero in the definition.}
\begin{gather} \label{mt2x}
m_{T2}^2 = 
\begin{cases}
A_T + \sqrt{A_T^2 - m_1^2 m_2^2}, & \mathrm{balanced}, \nonumber \\
\mathrm{max} (m_1^2, m_2^2), & \mathrm{unbalanced},
\end{cases}
\end{gather}
where $A_T \equiv  \sqrt{p_{1T}^2 + m_1^2}  \sqrt{p_{2T}^2 + m_2^2}+ p_{1T} \cdot p_{2T}$ and $p_{1,2T}$ and $m_{1,2}$ denote the transverse momenta and invariant masses of the visible systems in each decay.

This expression is manifestly invariant under independent longitudinal boosts of each parent, along with boosts that correspond to first boosting both decays to be purely transverse and subsequently applying equal, but opposite transverse boosts to each parent \cite{Cho:2007dh}.\footnote{We stress that this invariance group, large though it is, does not contain all back-to-back boosts of the parents.}

Just like $m_T$, the distribution of $m_{T2}$ has an upper endpoint encoding information about the masses of the particles involved.
However, the endpoint behaviour is complicated, if there are multiple visible particles in one or other decay, by the fact that either the unbalanced or balanced configuration may dominate, since both can reach the same maximum, $M$.
For balanced configurations, the maximum is reached when the invisible particles are parallel and transverse, leading to the same power-law behaviour as in (\ref{univ});
for unbalanced configurations, we need to maximize one or other visible invariant mass, leading to the fall-off given in (\ref{mv}), but with $m_V \to m_{T2}$
and with $n$ replaced by the number of invisible particles in one or other decay.
Now, since the endpoints are the same, the distribution there will be dominated by the component with the smaller power. For pair decays that are symmetric in the sense of having the same number of invisible particles on each side, the balanced solution dominates; for asymmetric decays, the unbalanced solution always dominates (provided that the decay with fewer invisible particles has multiple visible particles). We note that in the asymmetric case, the power is always an integer, whilst it is strictly a half-integer in the symmetric case. Thus, no ambiguity can arise between the two.
Just as for $m_T$ these results do not depend on the precise number of visible particles. 

\begin{figure}[t]
  \psfrag{2}{\color{MMAred}\boldmath$\left(\!1,\!1\!\right)$}
  \psfrag{3}{\color{MMAgreen}\boldmath$\left(\!1,\!2\!\right)$}
  \psfrag{4a}[br]{\color{MMAorange}\boldmath$\left(\!2,\!2\!\right)$}
  \psfrag{4b}[br]{\color{MMAorange}\boldmath$\left(\!1,\!3\!\right)$}
  \begin{center} 
    \includegraphics[width=0.99\linewidth]{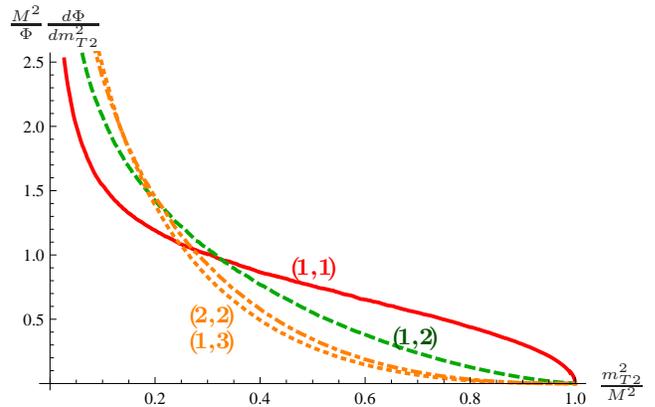}
    \end{center}
\caption{Normalized, phase-space distribution of $m_{T2}^2$ for $(k,l) \in \{ (1,1), (1,2), (2,2), (1,3) \}$ for  the process $2M \to 2P+(k+l)X$. \label{fig:mt3}} 
\end{figure}

In the simplest case of a single, massless, visible particle in each decay, $m_{T2}$ reduces to
\begin{gather} \label{mt2}
m_{T2}^2 = |p_{1T}||p_{2T}| + p_{1T} \cdot  p_{2T},
\end{gather}
which allows one to obtain an analytic form for the $m_{T2}$ distribution for two parents produced at rest in the lab or boosted as described above, given in (\ref{eqn:mt2n14}) for the case of two parents of equal mass, each decaying to $k$ or $l$, massless, invisible daughter particles. The distributions for the first few values of $n= k + l$ are shown in Fig.~\ref{fig:mt3}.
%%%%%%%%%%%%%%%%%%%%%%%%%%%%%%%%%%%%%%%%%%%%%%
%%%%%%%%%%%%%%%%%%%%%%%%%%%%%%%%%%%%%%%%%%%%%%
\subsection{Counting the number of parents}
%%%%%%%%%%%%%%%%%%%%%%%%%%%%%%%%%%%%%%%%%%%%%%
At this point, the reader might worry how a conscientious experimentalist, blessed with a sample of signal events, is to decide whether they came from decays of singly- or pair-produced parent particles and thus whether to try to count the invisible daughter particles using the $m_T$ or $m_{T2}$ distribution. We now suggest two possible ways to proceed.

The first possibility is simply to make an educated guess. For example, if the final state contains only a single visible particle, then single production seems likely; alternatively, if the final state contains two copies of all particles (or particle-antiparticle pairs), then pair production might reasonably be assumed. 

Perhaps the ideal strategy would be to try to independently count the number of parents and the number of invisible daughters. We have argued above that one expects that the distributions computed here will only give a good fit when one chooses the ``right'' observable for the process. If one chooses the wrong observable, then the absence of the desired invariance properties will lead to an observed distribution which is rather poorly fitted by the idealized distributions we present here. One thus might hope that one can identify the right observable (and in doing so, count the parents) by comparing multiple observables and seeing for which of those one obtains an acceptable fit. As an example, for single production followed by decay to two visible particles and one invisible particle, one would expect a good fit to the $m_T$ distribution, but a poorer fit to the $m_{T2}$ distribution.
%%%%%%%%%%%%%%%%%%%%%%%%%%%%%%%%%%%%%%%%%%%%%%
\section{Heavy invisible particles \label{sec:mass}}
%%%%%%%%%%%%%%%%%%%%%%%%%%%%%%%%%%%%%%%%%%%%%%
Thus far we have assumed that the invisible particles have negligible masses; we now redress this. 

Changing to a non-zero mass has several ramifications. 
The first issue arises in the definition of the observables $m_T$ and $m_{T2}$. For massless particles, we saw that the natural definition was to replace $m_I$ in (\ref{naive}) by its minimal value, namely zero. For massive particles, the obvious analogue would be to replace $m_I$ by its minimum value, namely the sum of the masses of the invisible particles present. Unfortunately, we do not expect that this quantity would have been measured at the time that the analysis proposed here could be carried out.\footnote{It is, by now, clear that there is enough information in kinematics (or equivalently $m_{T}$ or $m_{T2}$) to measure the DM mass at the LHC \cite{Barr:2010zj} in principle; it is also clear that the measurement will be extremely challenging in practice. We therefore take the view that the DM mass will be unknown at the time that efforts to count the number of invisible particles begin.} Again, the obvious solution, since $m_T$ as defined in (\ref{naive}) is a monotonically increasing function of $m_I$, is to replace $m_I$ by its minimal possible value, namely zero. 
Thus we retain the definition (\ref{mt}). This will lead, however, to yet further subtleties, as we shall see in the next Subsection.

A second issue is that the phase space available, in the limit that the invariant mass of the invisible particles approaches its minimum, differs depending on whether those invisible particles are massive or not. This should hardly come as a surprise, since the condition to be at the minimum differs in the two cases: in the massless case, the momenta of invisible particles must be parallel, whereas in the massive case, the momenta should, in addition, vanish, in the CM frame of the invisibles. There is, thus, less phase space available near threshold in the massive case; the explicit behaviours are given in (\ref{eqn:phs}) and (\ref{eqn:phs2}). As a result, (\ref{univ}) is changed in
the massive case, for a point-like decay of a single parent particle into a single, massless, visible particle and $n$ invisible particles, to
\begin{gather} \label{univmass}
\frac{d\Phi_{n+1}}{dm_T^2} \propto \left(1- \frac{m_{T}^2}{M_1^{2}}\right)^{\frac{3n}{2}-2}, 
\end{gather}
where now the endpoint is at $M_1 \equiv \frac{M^2 - \mu_n^2}{M}$ and where $\mu_n \equiv \Sigma m_X$ is the total mass of the invisible particles. We note that the behaviour is unchanged compared to the massless case for $n=1$.

Unlike the case of massless invisibles, we do not obtain the same formula for the $m_{T2}$ distribution for pair-produced particles. There, to be at the maximum of $m_{T2}$, the invisible particles produced in one decay must be at rest relative to one another, but need only be aligned (transversely) with the invisible particles produced in the other decay. Consequently, for point-like decays of a pair of parents, each into a single, massless, visible particle, we find
\begin{gather} \label{univmass1}
\frac{d\Phi_{k+1}\Phi_{l+1}}{dm_{T2}^2} \propto \left(1- \frac{m_{T2}^2}{M_2^{2}}\right)^{\frac{3n-5}{2}}, 
\end{gather}
where now the endpoint is given by $M_2^{ 2} \equiv \frac{(M^2 - \mu_k^2)(M^2 - \mu_l^2)}{M^2}$ and $n=k+l$.\footnote{We stress that this result is derived for parents produced at rest, or boosted as described after (\ref{mt2}). If one includes arbitrary back-to-back boosts, the endpoint shifts to $M^2 - \mathrm{min} (\mu_k, \mu_l)$ \cite{Barr:2009jv}, but the change in the endpoint behaviour is observed in simulations to be not greatly changed.}
Again, the behaviour is unchanged, compared to the massless invisible case, for $k=l=1$.

These changes in the power law in going from massless to massive invisible particles may seem confusing when one considers that (\ref{univmass}), for example, holds for all non-zero values of $\mu_n$, however small. Na\"{\i}vely, this suggests that there is a discontinuous (and observable) difference in the $m_T$ distributions at $\mu_n=0$ and as $\mu_n\rightarrow 0$. What really happens is that the behaviour in (\ref{univmass})
is obtained only in the region of the distribution where the invisible particle mass cannot be neglected in comparison with the distance from the endpoint. Further away, a transition to the behaviour in (\ref{univ}) occurs. The location of the transition thus depends on the mass itself and the distributions show a smooth behaviour as one takes the limit in which the invisible masses vanish. 
We learn that there is no paradox, but we also learn that the endpoint behaviour we can expect to see is contingent upon the unknown mass of the DM. Our recommended strategy for measuring $n$ will be to fit $n$ in the endpoint region of the distribution, treating the unknown invisible mass (or masses) as a nuisance parameter in the fit. This is unlikely to result in a precise determination of the invisible particle mass, but it should give us enough flexibility to reliably fit $n$.
%%%%%%%%%%%%%%%%%%%%%%%%%%%%%%%%%%%%%%%%%%%%%%
\subsection{Multiple visible particles\label{sec:massvis}}
%%%%%%%%%%%%%%%%%%%%%%%%%%%%%%%%%%%%%%%%%%%%%%
A third issue in the case of massive invisibles is that, with $m_T$ defined as in (\ref{mt}), the condition to be at the maximum (and indeed its location and endpoint behaviour) changes, if there are multiple visible particles. To see this, let us revisit the arguments of \S\ref{sec:single}. Equation (\ref{mtM}) is unchanged,
but since $m_I$ cannot vanish, being bounded below by the sum of the masses of the invisible particles, necessary conditions for $m_T$ to be at its maximum are that
$e_V q_I - q_V e_I = 0$, as before, and for $m_I$ to be at its minimum, $\tilde{m}_I$. Then, (\ref{mtM}) reduces to
\begin{multline}
m_{T}^2 = M^2 - \tilde{m}_I^2 -  2 e_V\left(\sqrt{\slashed{p}_T^2+ \tilde{m}_I^2}-\sqrt{\slashed{p}_T^2}\right).
\end{multline}
The expression in parentheses never vanishes and so to be at the maximum of $m_T$ one must also be at the minimum of $e_V$ and ergo $m_V$. Thus, to be at the maximum of $m_T$, one must also line up all the visible particles (if they are massless), or make their momenta vanish in their CM frame (if they are massive).\footnote{This is one of two effects that enables the (sum of) invisible particle mass(es) to be measured: if, as is the case here, the hypothesized mass in the definition of $m_T$ is smaller than the real mass, the maximum of $m_T$ is at the minimum of $m_V$, while if it is greater, the maximum of $m_T$ is at the maximum of $m_V$, leading to a kink in the endpoint of the $m_T$ distribution as a function of the hypothesized mass \cite{Cho:2007qv}. The other effect needs only one visible particle and relies on the varying alignment of the visible particle(s) with respect to upstream transverse momentum to generate a kink \cite{Gripaios:2007is}.}

\begin{figure}[t]
  \psfrag{0}{\color{MMAblue}\boldmath$x\!\!=\!\!0$}
  \psfrag{0.01}[br]{\color{MMAred}\boldmath$x\!\!=\!\!0.01$}
  \psfrag{0.1}[br]{\color{MMAgreen}\boldmath$x\!\!=\!\!0.1$}
  \psfrag{0.3}[br]{\color{MMAorange}\boldmath$x\!\!=\!\!0.3$}
  \begin{center} 
    \includegraphics[width=0.99\linewidth]{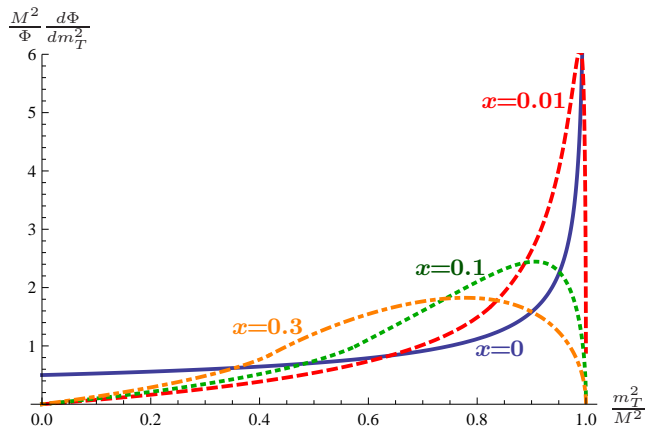}
    \end{center}
\caption{Phase-space distribution of $m_{T}$ for the process $2M \to 2P+1X$, for a massive invisible particle of mass $m_X$ for various values of the mass ratio $x=m_X^2/M^2$. \label{fig:mt4}} 
\end{figure}

To illustrate this, we give in  (\ref{eqn:massive2+1}) the $m_T$ distribution for
decay of a single parent to two massless visibles and a single massive
invisible. The reader may
verify that the power-law fall-off near the endpoint has changed from
minus one-half (as in the case of a massless invisible) to plus one-half (as in
the case of a massive invisible), because we now have to line up two (massless, visible) particles. Nevertheless, as illustrated in Fig.~\ref{fig:mt4}, the massless and massive distributions map onto each other as the invisible mass is sent to zero. 

Finally, we remark that this change in the endpoint behaviour, while dramatic, can easily be accommodated in performing the measurement of $n$, since the number of visible particles is easy to measure.
%%%%%%%%%%%%%%%%%%%%%%%%%%%%%%%%%%%%%%%%%%%%%%
%%%%%%%%%%%%%%%%%%%%%%%%%%%%%%%%%%%%%%%%%%%%%%

Further subtleties arise for $m_{T2}$, once we allow for multiple visible particles. Most importantly, either the endpoint at $\frac{(M^2 - \mu_k^2)(M^2 - \mu_l^2)}{M^2}$, corresponding to balanced configurations, or the one at $\mathrm{max} (M-\mu_{k,l})^2$, corresponding to unbalanced configurations, may be the true maximum in asymmetric decays. In the latter case, the endpoint behaviour of the $m_{T2}^2$ distribution is the same as that of the visible invariant mass distribution for the relevant decay, whose form we now discuss.
%%%%%%%%%%%%%%%%%%%%%%%%%%%%%%%%%%%%%%%%%%%%%%
\subsection{Invariant mass observables \label{sec:invmassmassive}}
%%%%%%%%%%%%%%%%%%%%%%%%%%%%%%%%%%%%%%%%%%%%%%
It is also worthwhile to consider the effect of massive invisible particles on the distributions of invariant masses of visible particles.
Again, let us consider a single particle of mass $M$ decaying into $l$ visible particles and $n$ invisible particles, all with arbitrary (but non-vanishing) masses. 
Just as in the case considered before, where all the invisible particles were massless, one may show that the condition to be at the maximum of the invariant mass of the $l$ visible particles is that all invisible particles be produced with zero momentum in the rest frame of the decaying particle. 
An explicit computation shows that the endpoint behaviour is given, for point-like decays, by \cite{BK}
\begin{gather}
\frac{d\Phi_{n+l}}{dm_V^2} \propto \left(1- \frac{m_V^2}{M_3^{ 2}}\right)^{\frac{3n}{2}-1},
\end{gather}
where now the endpoint is at $M_3 \equiv M - \mu_n$.
The coefficient of $n$ in the power-law fall-off is easy to understand. Here, unlike the massless case, if one adds an extra invisible particle, then the extra condition to be satisfied by that particle to be at the maximum of either $m_T$ or $m_V$ is the same: the extra particle must be at rest with respect to the invisible particles already present. Note however that the power laws are different in the massless and massive cases ($2n-1$ and $\frac{3n}{2}-1$ respectively), even though the condition to be at the maximum (all invisible particles' momenta vanishing) is the same in both cases.
%%%%%%%%%%%%%%%%%%%%%%%%%%%%%%%%%%%%%%%%%%%%%%
\subsection{Topology dependence \label{sec:masstop}}
%%%%%%%%%%%%%%%%%%%%%%%%%%%%%%%%%%%%%%%%%%%%%%
All the results obtained thus far for massive invisible particles were derived assuming point-like decays. In the limiting case of massless invisible particles, we saw that the endpoint behaviour of the transverse mass variable was independent of the decay topology. We now show, by means of a counter-example, that this is no longer true in the case of massive invisible particles. To wit, consider a single particle decaying into two invisible particles and a single visible particle. The behaviour of the transverse mass near its maximum is determined in part (as we saw in \S \ref{sec:massvis}) by the behaviour of the invariant mass of the invisible system near its minimum. For a point-like decay, the differential phase space in the square of the invisible invariant mass vanishes as the square root near its minimum, whereas it is flat if an on-shell intermediate lies between the two invisible particles. Happily, we would still fit the correct number of invisible particles, namely two, if we incorrectly assumed a point-like topology. We would, however, obtain an incorrect value of the mass, namely zero.

Thus, when invisible particle masses are non-negligible, both transverse and invariant mass distributions may exhibit a dependence on the invisible particle mass (or masses) and on the decay topology. As such, we recommend that, since neither of these is known {\em a priori}, they both be included as nuisance variables in the fit. 
Of course, one may hope that, in many cases, the dependence on these nuisance parameters will be small and will not result in an ambiguity in extracting the number of invisibles. We shall see instances of this in \S \ref{sec:examples}. As an example of how this may arise, the fact that transverse mass distributions are independent of the topology in the massless limit suggests that the dependence on the topology will remain small as long as the masses are not too large. 
%%%%%%%%%%%%%%%%%%%%%%%%%%%%%%%%%%%%%%%%%%%%%%
\section{Other effects \label{sec:other}}
%%%%%%%%%%%%%%%%%%%%%%%%%%%%%%%%%%%%%%%%%%%%%%
%%%%%%%%%%%%%%%%%%%%%%%%%%%%%%%%%%%%%%%%%%%%%%
\subsection{Matrix element and spin dependence \label{sec:matrix}}
%%%%%%%%%%%%%%%%%%%%%%%%%%%%%%%%%%%%%%%%%%%%%%
We now turn to other effects. The first of these is the modulation of phase space distributions by matrix-elements, due to spin or other effects.
By fitting the phase space distribution, we are implicitly assuming that these effects may be neglected. In this respect, our approach is reminiscent of the on-shell effective theory (OSET) approach \cite{ArkaniHamed:2007fw,Alves:2011wf}.

 It is easy to see, however, that there are cases in which these effects cannot be neglected for our purposes, in that they can, in principle, fake the effect of changing the number of invisible particles. Imagine, for example, that angular momentum considerations force two invisible particles to be always produced with equal momenta (or parallel if massless). Then, it is clear that the resulting distributions will be indistinguishable from the distributions that would have resulted from a single invisible particle (with mass given by the sum of the two individual masses). However, such effects appear only when decays occur at threshold and even then will be washed out by finite width effects. As an explicit example, in the decay $h\rightarrow 2W \rightarrow 2l 2\nu$ of a Higgs boson of mass $2m_W$, the two neutrinos are parallel in the limit of a narrow-width Higgs, but the $m_T$ distribution still shows the $n=2$ endpoint behaviour (\ref{univ}) in practice \cite{Barr:2009mx}.

A related worry is that the inclusion of spin effects will lead to two distributions, corresponding to different values of $n$, that are indistinguishable near their endpoints, if not elsewhere. Again, it is easy to construct an example where this could happen. Consider, following \cite{Csaki:2007xm}, a scalar particle, $A$, that
decays to an invisible fermion, $\psi$, and an off-shell, Dirac fermion, $\Psi$, which in turn decays to an invisible antifermion $\overline{\psi}$, and a visible scalar, $B$. Let the relevant Lagrangian couplings be
$ A \Psi P_L \psi + B \Psi P_R \psi + \mathrm{h.c.}$, where $2P_{L,R} \equiv 1 \pm \gamma_5$. Then, conservation of angular momentum forces the matrix element to vanish when the two invisible particles are parallel, which, we recall, is a necessary condition to be at the maximum of the $m_T$ distribution. As a result, the observed fall-off in the $m_T$ distribution will result in a measured value of $n>2$, if we try to fit the endpoint behaviour to (\ref{univ}).

As is well known (see e.g. \cite{Barr:2004ze}), similar effects may also arise if we consider invariant mass distributions of neighbouring visible particles in cascade decays. Such behaviour arises because the invariant mass is a function of the angle between the two visible particles, as measured in the rest frame of the intermediate particle; a polarization of the intermediate particle imparted by the first decay can then result in an angular dependence of the products of the second decay. Our expectation is that such effects will be larger for invariant mass observables than for $m_{T}$ or $m_{T2}$. Indeed, we have already seen in \S \ref{sec:massvis} that, in the case of massive invisible particles, one must minimize both $m_I$ and $m_V$ in order to be at the maximum of $m_T$. Thus, in order for spin effects to pollute the endpoint behaviour, they must appear in both the visible and invisible particle systems and they must not compensate each other.

Finally, it is possible that significant matrix-element effects will arise in another way. We have already seen that the transverse mass is maximized when the decay particles are produced perpendicular to the beam direction. One can imagine that such configurations will be preferred or disfavoured by the hard production process. An obvious example occurs in Drell-Yan production of $W$-boson, whose polarization disfavours larger values of $m_T$  \cite{Barger:1983wf}  (though the power-law fall-off is unchanged).
However, we expect that significant effects of this kind are unlikely to arise at the LHC, where several hard processes typically contribute to a given final state.

There is little more that can be done about these kinds of pathology at the general level, other than to be aware of their existence and to note that instances of them are rather rare (necessary conditions to generate effects in cascade decays, for example, are given in \cite{Wang:2006hk} and are seen to be rather stringent).
Moreover, even if such effects are present, they are typically small.
 As such, we are inclined to take the view that one should be delighted rather than disappointed if one is forced to contend with such a pathology in practice, since it will also give us the opportunity to make inferences about the matrix elements for processes involving new particles. The best way to proceed in such a case would seem to be a painstaking comparison of explicit Lagrangian hypotheses (with, {\em e. g.} differing spins, couplings and numbers of invisible particles) with data.

Finally, another special case is the one in which invisible particles do not have a single mass value, but rather form a nearly continuous spectrum of masses. Obvious examples are theories with large extra dimensions and the fundamental gravity scale close to a TeV, or unparticle theories possessing a conformal sector.
Both of these predict signals in e.g. monojet plus missing energy channels, which we might mistakenly interpret as single production of a heavy coloured particle which then decays to a single visible particle plus $n$ invisible particles.
We have checked that it should be relatively easy to distinguish these cases.  An obvious difference is that the graviton/unparticle distributions do not feature an endpoint and have an extremely rapid fall-off in $m_T$. They are therefore unlikely to be confused.
%%%%%%%%%%%%%%%%%%%%%%%%%%%%%%%%%%%%%%
%%%%%%%%%%%%%%%%%%%%%%%%%%%%%%%%%%%%%%%%%%%%%%
\subsection{Finite width effects \label{sec:width}}
%%%%%%%%%%%%%%%%%%%%%%%%%%%%%%%%%%%%%%%%%%%%%%
A decaying particle inevitably has a finite width, given by a Breit-Wigner distribution. This has two important effects. Firstly, the distributions above (derived for fixed mass and zero width) must be convolved with a Breit-Wigner distribution. Since the latter distribution is non-vanishing for arbitrarily large masses, the singular endpoints discussed above disappear, to be replaced by distributions which fall sharply beyond the peak mass. To illustrate this, consider convolving the phase-space $m_T$ distribution for a two-body decay (\ref{eqn:mtdist11}) with the Breit-Wigner distribution
$\left((m_T^2 - M^2)^2 + \Gamma^2 M^2\right)^{-1}$; the result is
\begin{gather}
\frac{d \Phi_2}{d m_T^2} \propto \frac{\cos \theta}{\left((1-\frac{m_T^2}{M^2})^2 + \frac{\Gamma^2}{M^2}\right)^{\frac{1}{4}}},
\end{gather}
where $\tan 2\theta \equiv \frac{\Gamma M}{M^2 - m_T^2}$.\footnote{The precise result depends on what one means by ``the phase space $m_T$ distribution''. Here we have used $\frac{1}{\sqrt{M^2 - m_T^2}}$, which corresponds to the decay rate under the assumption that all kinematic and dynamic dimensionful parameters are set by the variable parent mass and are distributed according to the Breit-Wigner formula.}  

\begin{figure}[t]
  \psfrag{1 }{\color{MMAblue}\boldmath$n\!\!=\!\!1$}
  \psfrag{0b}{\color{MMAblue}\boldmath$y\!\!=\!\!0$}
  \psfrag{0.1b}{\color{MMAblue}\boldmath$y\!\!=\!\!0.1$}
  \psfrag{2 }{\color{MMAred}\boldmath$n\!\!=\!\!2$}
  \psfrag{0r}{\color{MMAred}\boldmath$y\!\!=\!\!0$}
  \psfrag{0.1r}[br]{\color{MMAred}\boldmath$y\!\!=\!\!0.1$}
  \begin{center} 
\includegraphics[width=0.99\linewidth]{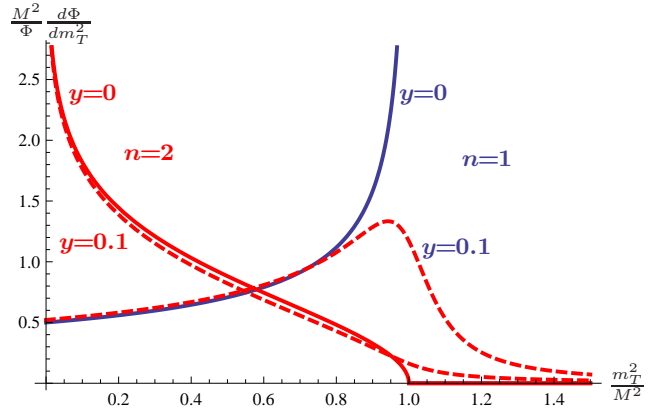}
    \end{center}
\caption{Phase space distribution of $m_{T}$ for the process $M \to
  P+nX$, with fractional width $y=\Gamma/M$, for $n=1,\;2$. \label{fig:width}} 
\end{figure}

We plot the distribution for various values of the fractional width $\Gamma/M$, in Fig.~\ref{fig:width}, together with the corresponding distribution for two invisible particles.
The presence of finite-width effects of this type means that it makes no sense to try to count invisible particles by directly fitting empirical distributions to phase space. Rather, we should fit the phase space distribution (with the normalization unfixed and the invisible mass kept as a free, nuisance parameter), convolved with a Breit-Wigner distribution (whose fractional width is also, in general, a nuisance parameter), in the region of the endpoint. 
%%%%%%%%%%%%%%%%%%%%%%%%%%%%%%%%%%%%%%%%%%%%%%
%%%%%%%%%%%%%%%%%%%%%%%%%%%%%%%%%%%%%%%%%%%%%%
%%%%%%%%%%%%%%%%%%%%%%%%%%%%%%%%%%%%%%%%%%%%%%
\subsection{Backgrounds}
%%%%%%%%%%%%%%%%%%%%%%%%%%%%%%%%%%%%%%%%%%%%%%
It was argued in \cite{Barr:2009wu} that the transverse mass $m_T$ (and its cousin $m_{T2}$) also has the desirable property that it provides a good discrimination between signal and background events, at least if the new physics signal is sufficiently heavy.\footnote{More precisely, the mass difference between the parents and the invisible daughter in the new physics signal should be large compared to mass scales associated with Standard Model backgrounds.}  To wit, whilst the signal can have large values of $m_T$, background events have small values of $m_T$. This property is not only useful for discovering new physics (and indeed the $m_{T2}$ distribution in the dijets plus missing energy channel using the 2010 LHC data gave the strongest constraint on squark masses \cite{daCosta:2011qk,Lester:2011nj}), but it also tells us that the upper endpoint of the $m_T$ or $m_{T2}$ distributions, where we propose to carry out the measurement of $n$, will have a suppressed background contamination.
%%%%%%%%%%%%%%%%%%%%%%%%%%%%%%%%%%%%%%%%%%%%%%
\subsection{Upstream transverse momentum}
%%%%%%%%%%%%%%%%%%%%%%%%%%%%%%%%%%%%%%%%%%%%%%
The $m_T$ and $m_{T2}$ variables (and, {\em ergo}, their distributions) are both invariant under longitudinal boosts
of the rest frame of the decaying parent or parents. The $m_{T2}$ distributions in (\ref{eqn:mt2n14}) are, furthermore, invariant under a larger set of boosts as described in \S\ref{sec:pair}.
Neither variable is invariant under overall, transverse boosts of the (pair of) parent(s), however, and we now proceed to consider the effect of these.\footnote{For decays involving jets in the final state, radiation from the initial state poses an additional combinatorial ambiguity. Methods to reduce this using the $m_{T2}$ variable were discussed in \cite{Alwall:2009zu}.} An additional benefit of the original definition of $m_T$ \cite{vanNeerven:1982mz} as used to measure the mass of the $W$-boson \cite{Arnison:1983rp,Banner:1983jy} is that there is no shift in $m_T$ under a transverse boost $\beta_T$, at linear order in $\beta_T$ \cite{Barger:1983wf}. 
Any shift thus goes as $\beta_T^2$ and is small for production of heavy states near threshold. What is more, since the endpoint of the $m_T$ distribution is invariant under Lorentz boosts, any shift in the $m_T$ distribution must also vanish as some power of the distance from the endpoint. There is thus a double suppression near to the endpoint.

Unfortunately, neither of these results holds generally when we extend $m_T$ to massive final state particles (visible or invisible). Indeed, one may easily show that the shift in $m_T$ (as defined in (\ref{naive})) under a transverse boost vanishes at linear order only if the correct invisible mass is used in the definition of $m_T$ and only if $m_I = m_V$. Similarly, it is easy to show that the upper endpoint of the $m_T$ distribution is Lorentz invariant only if the correct invisible mass is used in the definition of $m_T$.\footnote{Again, it is this phenomenon that is responsible for the kink in the endpoint of the $m_T$ distribution that allows the masses to be measured with just one visible particle \cite{Gripaios:2007is}.}
Since we do not know the invisible mass {\em a priori}, this cannot be done.

In conclusion, the distributions considered here will be sensitive to the presence of upstream transverse momentum in the event, in whatever form. Again, we hope the resulting distortions of distributions will not be so severe as to lead to a mismeasurement of the number of invisible particles, but caution should be exercised. Of course, if one did have confidence in one's understanding of the spectrum of upstream transverse momentum associated with a new physics signal, then one could compensate for its effects.
%%%%%%%%%%%%%%%%%%%%%%%%%%%%%%%%%%%%%%%%%%%%%%
\section{Examples \label{sec:examples}}
%%%%%%%%%%%%%%%%%%%%%%%%%%%%%%%%%%%%%%%%%%%%%%
We now apply our fitting procedure to various examples including both massless and massive invisibles.  These test cases feature many of the effects discussed above and, as such, serve to verify the viability of our algorithm for counting the number of invisibles in the final state in the presence of, for instance, pdfs, cascade decays, finite widths and spin effects.

%%%%%%%%%%%%%%%%%%%%%%%%%%%%%%%%%%%%%%%%%%%%%%
\subsection{Light invisible particles \label{sec:eg1}}
%%%%%%%%%%%%%%%%%%%%%%%%%%%%%%%%%%%%%%%%%%%%%%
%%%%%%%%%%%%%%%%%%%%%%%%%%%%%%%%%%%%%%%%%%%%%%
\begin{figure*}[t!]
  \psfrag{csomw}{$m_W^2\frac{d N}{d m_T^2}$}
  \psfrag{mTomw}{$\frac{m_T^2}{m_W^2}$}
  \psfrag{csomh}{$m_h^2\frac{d N}{d m_T^2}$}
  \psfrag{mTomh}{$\frac{m_T^2}{m_h^2}$}
  \psfrag{cs2omw}{$m_W^2\frac{d N}{d m_{T2}^2}$}
  \psfrag{mT2omw}{$\frac{m_{T2}^2}{m_W^2}$}
  \psfrag{cs2omt}{$m_t^2\frac{d N}{d m_{T2}^2}$}
  \psfrag{mT2omt}{$\frac{m_{T2}^2}{m_t^2}$}
  \psfrag{1 }{\color{MMAblue}\boldmath$n\!\!=\!\!1$}
  \psfrag{2 }[rb]{\color{MMAred}\boldmath$n\!\!=\!\!2$}
  \psfrag{ 2}{\color{MMAblue}\boldmath$n\!\!=\!\!2$}
  \psfrag{3 }[rb]{\color{MMAred}\boldmath$n\!\!=\!\!3$}
\centering
\subfloat[$m_T^2$ distribution for $W$ production and leptonic decay at the Tevatron.]{\includegraphics[width=0.45\linewidth]{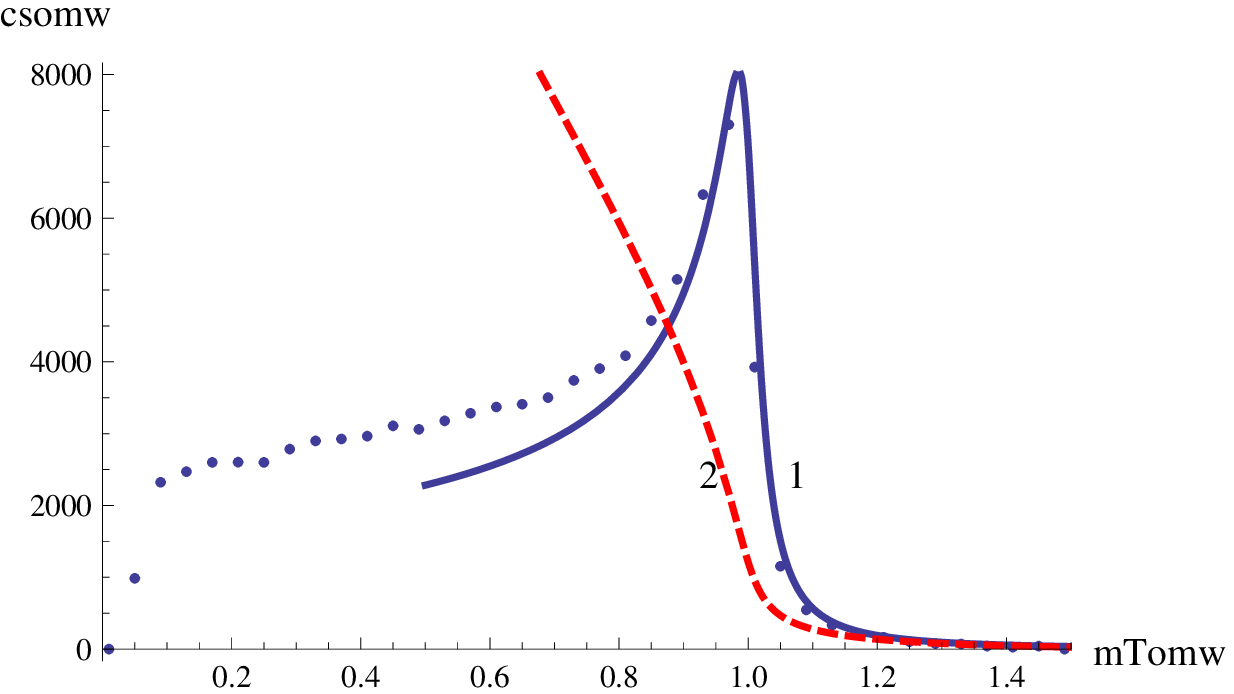}}\qquad
\subfloat[$m_T^2$ distribution for $h\rightarrow W^+ W^-$ and leptonic $W$ decay at the LHC, with $m_h=180$ GeV.]{\includegraphics[width=0.45\linewidth]{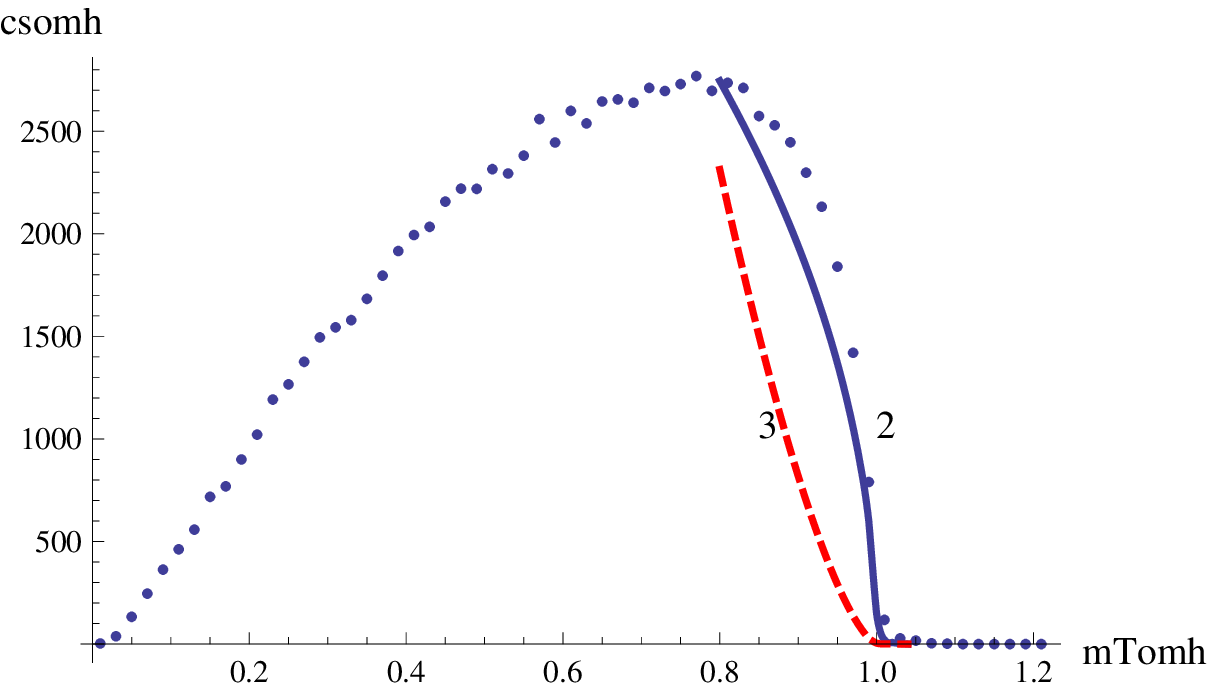}}\\
\subfloat[$m_{T2}^2$ distribution for $h\rightarrow W^+ W^-$ and leptonic $W$ decay at the LHC, with $m_h=180$ GeV.]{\includegraphics[width=0.45\linewidth]{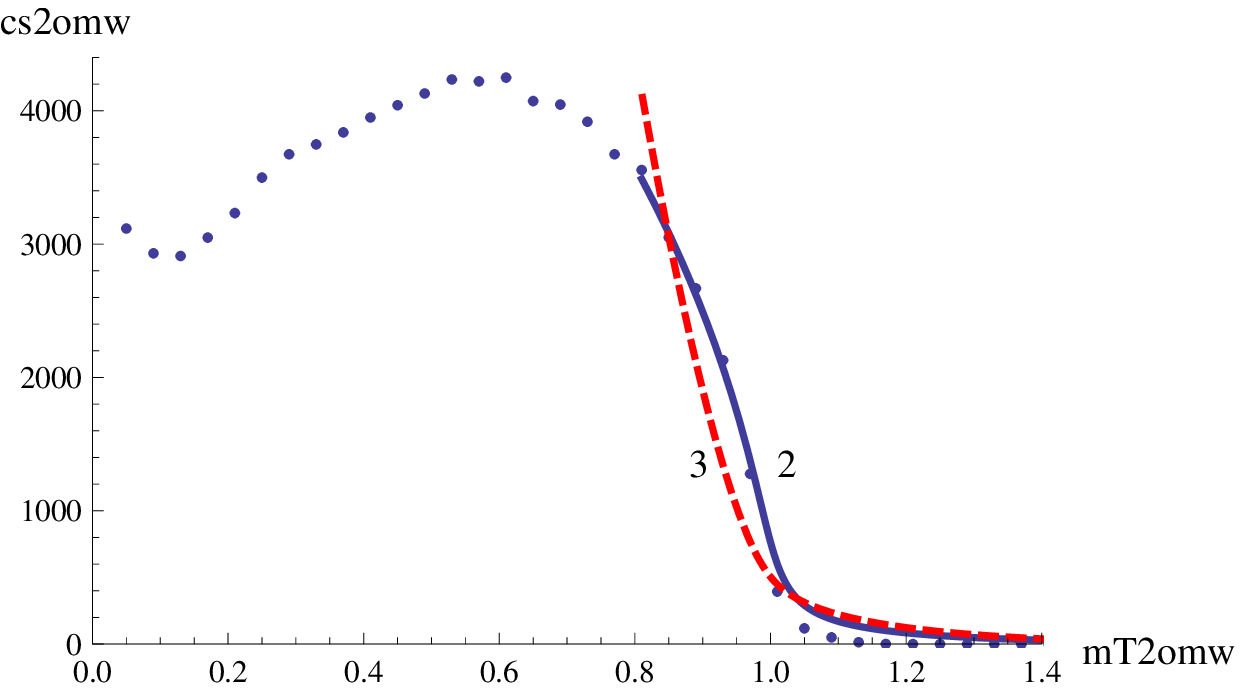}}\qquad
\subfloat[$m_{T2}^2$ distribution for $t\bar{t}$ production and leptonic decay at the LHC.]{\includegraphics[width=0.45\linewidth]{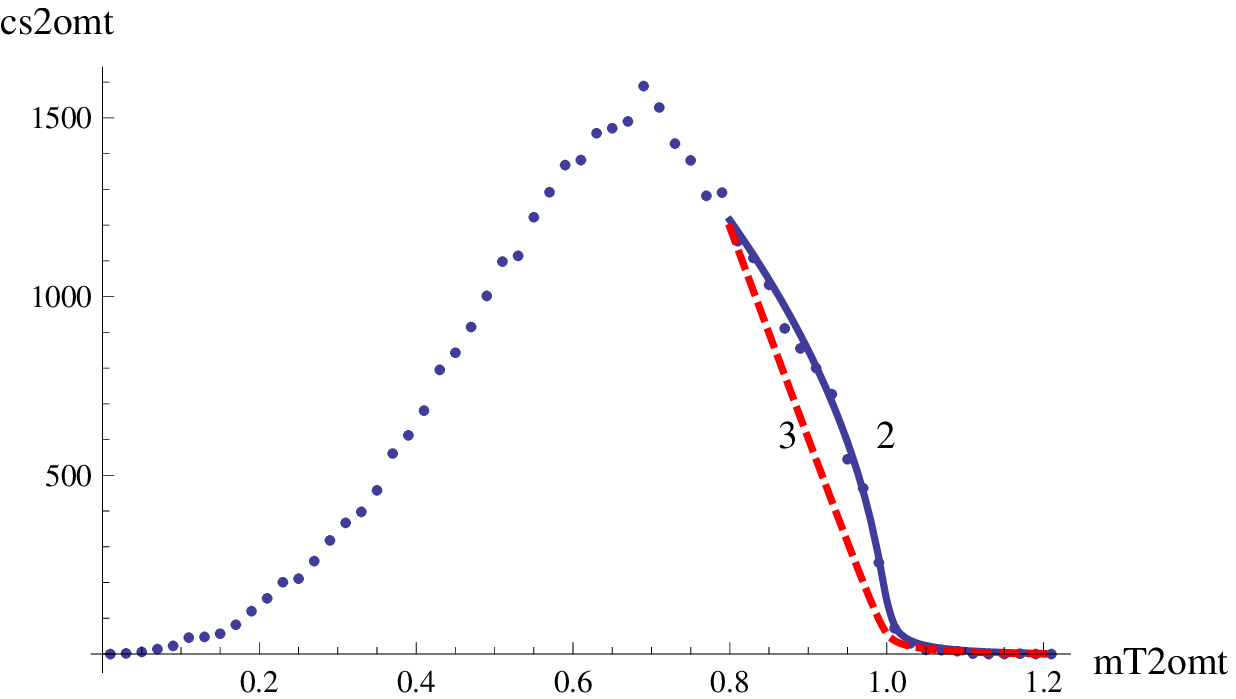}}
\caption{Standard Model examples with final state neutrinos, along with
  the best and next-best fits for distribution of transverse mass variables
  near endpoint.  The fits are to the relevant endpoint power
  law (see Table \ref{tab:laws}) convolved with a Breit-Wigner, with fractional width set
  to the true value.  Details of Monte Carlo event generation are given in the text. \label{fig:fits}}
\end{figure*}
%%%%%%%%%%%%%%%%%%%%%%%%%%%%%%%%%%%%%%%%%%%%%%
For realistic examples with massless invisibles, we need look no further than the Standard Model.  Specifically, we consider $W$-boson production and leptonic decay at the Tevatron; Higgs boson production and di-leptonic decay via a pair of intermediate $W$-bosons at the LHC, with $m_h=180$ GeV; and finally pair production of top quarks undergoing di-leptonic decay at the LHC.

We generate events in each case at tree level using {\tt MadGraph4/MadEvent}~\cite{Alwall:2007st}.  We analyze these at parton level, using
truth information to avoid combinatorial ambiguities, {\em e.g.} in pairing 
the $b$-jets with leptons in pairs of top quark decays.
 We then fit the transverse mass (or $m_{T2}$)
distributions in the region of the endpoint using the relevant
power-law behaviour for massless invisibles (summarized in Table \ref{tab:laws}),
convolved with a Breit-Wigner distribution, whose fractional width is fixed to the true value for simplicity.  The
resulting $R$-squared values, quantifying the goodness of fit, for
various $n$ are shown in Table \ref{tab:Rsq} and the best and
next-best fits are superimposed on the ``data'' in Fig. \ref{fig:fits}.
%%%%%%%%%%%%%%%%%%%%%%%%%%%%%%%%%%%%%%%%%%%%%%
%%%%%%%%%%%%%%%%%%%%%%%%%%%%%%%%%%%%%%%%%%%%%%
\begingroup
\squeezetable
\begin{table}
\begin{center}
\begin{ruledtabular}
\begin{tabular}{ccccc}
%\cline{3-5}
 & &  \multicolumn{3}{c}{$R$-squared}\\
\cline{3-5}
Process & Observable & $n=1$ & $n=2$ & $n=3$\\ 
\hline
$W\rightarrow\ell\nu$ &$m_T$ & 0.982 & 0.723 & 0.529 \\ %\hline
\multirow{2}{*}{$h\rightarrow W^+ W^-\rightarrow
 \ell^+\ell^-\nu\bar{\nu}$} & $m_T$ & 0.501 &
 0.977 & 0.779 \\ 
%\cline{2-5}
& $m_{T2}$ & 0.630 & 0.996 & 0.952 \\ %\hline
$t\bar{t}\rightarrow
 W^+W^-b\bar{b}\rightarrow\ell^+\ell^-\nu\bar{\nu}b\bar{b}$ & $m_{T2}$
 & 0.458 & 0.999 & 0.971\\ %\hline
\end{tabular}
\end{ruledtabular}
\end{center}
\caption{Goodness-of-fit values for various Standard Model processes
  with neutrinos,
  fitted with the relevant endpoint power law (see Table \ref{tab:laws}), convolved with a Breit-Wigner.  Fractional widths of parents
  are fixed to their true values. Best fits are shown in Fig. \ref{fig:fits}.\label{tab:Rsq}} 
\end{table}
%%%%%%%%%%%%%%%%%%%%%%%%%%%%%%%%%%%%%%%%%%%%%%

We see that our simple algorithm picks out the correct value of $n$ in each instance, although the $R$-squared values for $n=2$ and $n=3$ are sometimes close.  Furthermore, although the $n=2$ distribution is the best fit for the $m_T$ distribution in $h\rightarrow W^+W^-$, the quality of the fit is not quite as good as in the other cases and the actual fall-off near the endpoint is somewhat steeper than $n=2$ would warrant.  We believe this to be a hallmark of spin correlations: in the limit that $m_h\rightarrow 2m_W$ and the widths go to zero, conservation of angular momentum forces the two neutrinos to be emitted in the same direction, hence acting like a single invisible.\footnote{In fact we have verified that even for $m_h=2m_W$, the best fit still corresponds to the power law for $n=2$.}  We see a similar effect in $t\bar{t}$ production if we pretend that the charged leptons, rather than the neutrinos, are invisible.  This too is as expected, since spin correlations in the $t\bar{t}$ system have a stronger effect on the distribution of the charged leptons~\cite{Mahlon:2010gw}. 

Finally, we note that for $t\bar{t}$ production we obtain a good fit
using the endpoint power-law for point-like decays, even though the intermediate $W$-bosons are produced on-shell.
%%%%%%%%%%%%%%%%%%%%%%%%%%%%%%%%%%%%%%%%%%%%%%
\subsection{Heavy invisible particles \label{sec:eg2}}
%%%%%%%%%%%%%%%%%%%%%%%%%%%%%%%%%%%%%%%%%%%%%%
Processes with heavy invisibles in the final state are abundant in
theories going beyond the Standard Model.  We choose to fit an example where
both spin correlations and topology-dependence are known to play a r\^{o}le, namely the supersymmetric process of
$\tilde{u}_L$ pair production, followed by decays $\tilde{u}_L \to u_L \tilde{\chi}_2^0 \to u_L \ell_R^+ \tilde{\ell}_R^- \to u_L \ell_R^+ \ell_R^- \tilde{\chi}_1^0$
\cite{Barr:2004ze}. We use {\tt SOFTSUSY} \cite{Allanach:2001kg} to
compute the SUSY spectrum, generate the events using {\tt
  Madgraph/Madevent} and BRIDGE \cite{Meade:2007js} and analyse them
using truth information as before.  For the fits we use RAMBO
\cite{Kleiss:1985gy} to generate phase space for the point-like decays $M+M \rightarrow (3+3)P + (k+l)X$ with varying
invisible mass; these are again convolved with a Breit-Wigner of fixed, true fractional width. The
resulting $R$-squared values for
different $n=k+l$ and invisible mass are shown in Table \ref{tab:Rsqmassive}, with the best and
next-best fits superimposed on the ``data'' in
Fig. \ref{fig:fitmassive}.  Note that in spite of spin correlations
that significantly affect the invariant mass distributions for
pairs of visibles in this decay, once again the best fit for $m_{T2}$ is for the
correct value of $n$, and also for the invisible mass that is closest
to the input value of 115 GeV. Moreover, as in the $t\bar{t}$ case, point-like
phase space gives a very good fit to the distribution, even though the
actual process is a 3-step cascade decay.  Unfortunately it is
not possible to say definitively, using the $m_{T2}$ distribution alone, that this process is not instead
$n=3$ with light invisibles.  This is an ambiguity that comes about
because the endpoint behaviour of $m_{T2}$ here is governed by the
unbalanced solution which falls off linearly; it can, however, be
resolved by separately examining the visible invariant mass
distributions on each leg of the decay (see Fig
\ref{fig:fitmassive}(b)).\footnote{As expected, the best fits for
  $m_V^2$ are not as good as those for transverse masses, since we are
  fitting point-like phase space in spite of the known, large dependence of
  these distributions on topology.} Finally, one
might worry that the results of the fits might change if one allowed the fractional width in the
Breit-Wigner distribution to float.  We would argue that plausible values for these,
at least for small $n$, can be chosen by eye from the distributions,
and, moreover, that scanning over these is unlikely to make a
significant difference to the best fit.  (For example it is difficult to see how a peaked
$n=1$ distribution could be mistaken for $n=2$, even allowing for arbitrary fractional width.)

%%%%%%%%%%%%%%%%%%%%%%%%%%%%%%%%%%%%%%%%%%%%%%
\begingroup
\squeezetable
\begin{table}
\begin{center}
\begin{ruledtabular}
\begin{tabular}{c c c c c c c c}
%\cline{2-4}
  &  \multicolumn{3}{c}{$R$-squared $\left(m_{T2}^2\right)$} &
  \multicolumn{2}{c}{$R$-squared $\left(m_{V1}^2\right)$} & \multicolumn{2}{c}{$R$-squared $\left(m_{V2}^2\right)$}\\
\cline{2-8}
 $m_X/$ GeV & $n=2$ & $n=3$ & $n=4$ & $k=1$ & $k=2$ & $l=1$ & $l=2$\\ 
% \multicolumn{1}{|c|}{ $m_X$ (GeV)}& $n=2$ & $n=3$ & $n=4$\\ 
\hline
0 & 0.947 & 0.992 & 0.899 & 0.992 & 0.855 & 0.987 & 0.843\\
100 & 0.998 & 0.889 & 0.822 & 0.993 & 0.921 & 0.998 & 0.911\\
450 & 0.859 & 0.991 & 0.870 & 0.987 & 0.911 & 0.994 & 0.900\\
%\hline
\end{tabular}
\end{ruledtabular}
\end{center}
\caption{Goodness-of-fit values for the supersymmetric cascade decay in \cite{Barr:2004ze}, with neutralino mass of 115 GeV, fitted to 
the endpoint behaviour of generated $(3+k)+(3+l)$-body phase space, where $k+l=n$, with varying invisible mass and convolved with a Breit Wigner of fixed fractional width.  Best fits are shown in Fig. \ref{fig:fitmassive}.\label{tab:Rsqmassive}}
\end{table}
\endgroup
%%%%%%%%%%%%%%%%%%%%%%%%%%%%%%%%%%%%%%%%%%%%%%
%%%%%%%%%%%%%%%%%%%%%%%%%%%%%%%%%%%%%%%%%%%%%%
\begin{figure*}[t!]
  \psfrag{cs2omsq}{$M^2\frac{d N}{d m_{T2}^2}$}
  \psfrag{mt2omsq}{$\frac{m_{T2}^2} {M^2}$}
  \psfrag{2}{\color{MMAblue}\boldmath$n\!\!=\!\!2, m_X\!\! =\!\! 100$ GeV}
  \psfrag{3}[rb]{\color{MMAred}\boldmath$n\!\!=\!\!3, m_X \!\!=\!\! 0$}
  \psfrag{csv1omsq}{$M^2\frac{d N}{d m_{V}^2}$}
  \psfrag{mv1omsq}{$\frac{m_{V}^2} {M^2}$}
  \psfrag{m0}[rb]{\color{MMAblue}\boldmath$k\!\!=\!\!1, m_X\!\! =\!\! 0$}
  \psfrag{m100}{\color{MMAred}\boldmath$k\!\!=\!\!1, m_X \!\!=\!\!
  100$ GeV}
\centering
\subfloat[$m_T^2$ distribution for SUSY cascade process at the
  LHC.]{\includegraphics[width=0.45\linewidth]{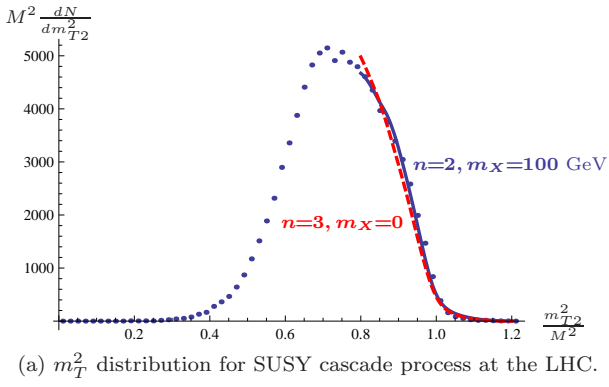}}\qquad
\subfloat[$m_V^2$ distribution for one leg of same SUSY cascade process at the
  LHC.  Similar results are obtained for the other leg.]{\includegraphics[width=0.45\linewidth]{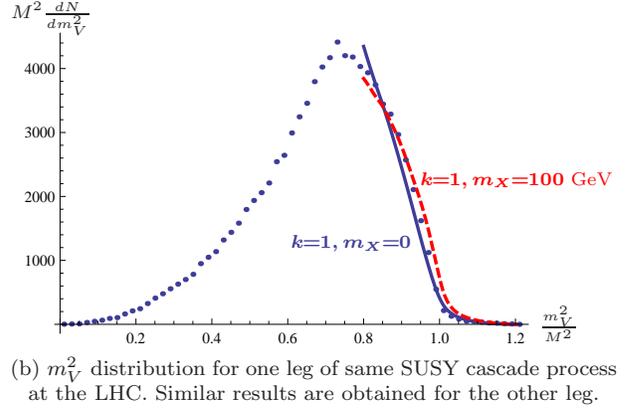}}
\caption{Best and next-best fits for the SUSY cascade process in \cite{Barr:2004ze} with 6 visible particles and a pair of 115 GeV neutralinos in the final state. The fits are to the endpoint behaviour of generated $(3+k)+(3+l)$-body phase space, where $k+l=n$, with varying invisible mass and convolved with a Breit Wigner of fixed fractional width.  Details of the Monte Carlo event generation are given in the text.\label{fig:fitmassive}}
\end{figure*}
%%%%%%%%%%%%%%%%%%%%%%%%%%%%%%%%%%%%%%%%%%%%%%

%%%%%%%%%%%%%%%%%%%%%%%%%%%%%%%%%%%%%%%%%%%%%%
\section{Discussion \label{sec:disc}}
%%%%%%%%%%%%%%%%%%%%%%%%%%%%%%%%%%%%%%%%%%%%%%
The existence of DM in the cosmos is now established beyond reasonable doubt; the prospect that we may imminently manufacture and be able to study DM in the laboratory is a tremendously exciting one, for particle- and astro-physicists alike. Here we have argued that a useful preliminary way to study an invisible DM candidate at the LHC would be to count the invisible particles produced in an event and we have outlined a procedure by which such a measurement  might be achieved. Doing so would not only increase one's confidence that one really had made DM (if multiple production could be established), but could also yield deep insights about the nature of DM and its stabilizing symmetry. 

To this end, we have shown that certain observables have a strong, power-law dependence, in their endpoint behaviour, on the number of invisible particles; our results are summarized in Table \ref{tab:laws}. This endpoint behaviour follows from simple kinematic considerations (namely that the invisible particles should be produced either parallel to each other or at rest) and is, to some extent, universal. 
The universality, together with the strength of the dependence, means that counting invisible particles using the endpoint behaviour should be robust with respect to a number of conceivable complications and gives us high hopes that one could reliably count the number of invisible particles in practice, given favourable conditions.
%%%%%%%%%%%%%%%%%%%%%%%%%%%%%%%%%%%%%%%%%%%%%%
\begingroup
\squeezetable
\begin{table}
\begin{center}
\begin{ruledtabular}
\begin{tabular}{l c  c  c  c}
%\hline
& & &\multicolumn{2}{c}{Exponent}  \\
\cline{4-5}
%\cmidrule(1){2-3}
Production & Observable & Invisibles & $\mu=0$  & $\mu\neq 0$ \\
\hline 
Single & $m_T$ & $n$ & $n - \frac{3}{2}$ & $\frac{3n}{2}-2$ \\
Symmetric pair & $m_{T2}$ & $n=k+l$& $k+l - \frac{3}{2}$ & $\frac{3(k+l)}{2}-\frac{5}{2}$ \\
Asymmetric pair & $m_{T2}$ & $n=k+l, k < l$ & $2k - 1$ & - \\
- & $m_{V}$& $n$ & $2n - 1$ & $\frac{3n}{2}-1$ \\
%\hline
\end{tabular}
\end{ruledtabular}
\end{center}
\caption{Endpoint power-law fall-offs for transverse- and invariant-mass distributions, for massless ($\mu = 0$) and massive  ($\mu \neq 0$) invisible particles. Valid for point-like decays  (arbitrary decay topologies for $m_{T}$ in the $\mu = 0$ case) and only for decays with a single visible particle 
for $m_T$ or $m_{T2}$ in the $\mu \neq 0$ case. The formula for asymmetric pair decays applies only if there are multiple visible particles in the decay with fewer invisibles; if not, the formula for symmetric pair decays applies. The $m_V$ observable may be employed for any production process, provided the visible decay products of a single parent can be isolated. \label{tab:laws}}
\end{table}
\endgroup
%%%%%%%%%%%%%%%%%%%%%%%%%%%%%%%%%%%%%%%%%%%

We have focused on transverse mass variables (and their cousins, such as $m_{T2}$) and invariant mass observables. Both feature maximal endpoints with a strong dependence on the number of invisible particles. The invariant mass observables have the advantages that they do not involve the missing energy, such that the empirical distributions may be cleaner, and simulations suggest that they have reduced sensitivity to invisible particle masses in cascade decays. In contrast, transverse mass observables have the advantage that they are less sensitive to the details of the decay topology (indeed, $m_T$ is independent of it, in the limit that invisible particle masses are negligible) and that we expect them to be less sensitive to spin effects.

There are important respects in which our results are not universal, which we have taken pains to point out. The most important of these is the effect of the mass of invisible particles.  This effect may be significant and may even lead to ambiguities in counting the number of particles. For example, Table \ref{tab:laws} shows that, for single production, one cannot distinguish $(n=3,\mu = 0)$ from $(n=4,\mu \neq 0)$ using the power-law fall-off of $m_V$ alone.
There is also a secondary dependence on the widths of the parents involved and, in general, on the topologies via which they decay. To compensate for this, we propose that experiments allow these variables to float (within reasonable ranges) when fitting the endpoint behaviour. We have argued, nevertheless, that there will often be a high degree of insensitivity, in which case  the poor constraint on the value of the nuisance parameter that results may provide a useful cross-check. We have also identified pathological cases, in which matrix-element effects may lead to ambiguities or errors in inferring the number of invisible particles.

In the massless invisible case, we have argued that it suffices to fit the endpoint power-law behaviour (convolved with finite-width and, eventually, detector effects). In the massive case, there is a transition between massive and massless power-law behaviours and so the approach needs to be modified. Concretely, our proposed strategy is as follows. Given some missing energy signal in a particular channel and associated transverse- or invariant-mass distributions,
generate phase-space distributions, corresponding to single or pair production of parents at rest in the lab frame, with the observed number of visible particles. This should be done for varying numbers of invisible particles and their masses (and, if necessary, for various decay topologies). Convolve these distributions with Breit-Wigner distributions (of varying width if necessary) and the appropriate detector response. Fit the resulting distributions to the signal in the endpoint region and extract the best fit value of the number of invisible particles.

Our proposal might be further developed in various ways. Firstly, we have focused our attention on only a limited set of observables and only on their maxima. More generally, many observables feature singular points in their distributions that result from the projection of phase space into its observable subspace \cite{Kim:2009si}; the behaviour near these generalized singularities should also contain information about the number of invisible particles and merits further exploration. Secondly, we have not dealt with the issue of combinatoric ambiguities arising in identical pair decays or from the presence of initial state radiation. Variables generalizing $m_{T2}$ that deal with these (and which enjoy similar boundedness properties) have been proposed \cite{Lester:2007fq,Alwall:2009zu} and it would be of interest to extend the study carried out here to them. Thirdly, we have shown that there can be a dependence on the topology in various cases and a fuller study of such effects would be desirable. Finally, even though we have yet to see evidence for new, invisible particles produced at the LHC, now would seem to be the ideal time for experiments to validate and refine our proposal, by counting the neutrinos which certainly have been abundantly produced in various SM processes, of which we have discussed various examples.   
%%%%%%%%%%%%%%%%%%%%%%%%%%%%%%%%%%%%%%%%%%%%%%

%%%%%%%%%%%%%%%%%%%%%%%%%%%%%%%%%%%%%%%%%%%%%%
\section*{Acknowledgments}
%%%%%%%%%%%%%%%%%%%%%%%%%%%%%%%%%%%%%%%%%%%%%%
We thank N. ~Arkani-Hamed, A.~Barr, R.~Franceschini, C.~Lester, and R.~Rattazzi for discussions. We also thank J.~Alwall, E. ~Dobson, R.~Frederix, M.~Reece and P. ~Skands for assistance with Monte Carlo event generation and fits.
%%%%%%%%%%%%%%%%%%%%%%%%%%%%%%%%%%%%%%%%%%%%%%
\appendix
%%%%%%%%%%%%%%%%%%%%%%%%%%%%%%%%%%%%%%%%%%%%%%
\section{Useful Formul\ae \label{sec:app}}
%%%%%%%%%%%%%%%%%%%%%%%%%%%%%%%%%%%%%%%%%%%%%%
The differential phase space for the decay of a particle $M$ into $n$ final-state particles $P_i$ is
\begin{multline}
d \Phi_{n} (M \to P_1 \dots P_n) = \\ \left( \Pi_{i=1}^n \frac{d^3 p_i}{(2\pi)^32E_i} \right)(2\pi)^4 \delta\left( E-\sum_{i=1}^n E_i\right) \delta^3\left( P-\sum_{i=1}^n p_i\right) .
\label{phase}
\end{multline}
Here $P$, $E$ and $p_i$, $E_i$ are the momenta and energies of the initial and final state particles, respectively. 
In the case in which all particles $P_i$ are massless, the integrated phase space is
\beq \label{eqn:phs}
\Phi_{n} (M \to P_1 \dots P_n) =\frac{M^{2(n-2)}}{(n-1)!(n-2)! \, 2^{4n-5} \, \pi^{2n-3}}.
\eeq
In the case in which all particles $P_i$ are massive, with masses $m_i$, the integrated phase space is given, in general, by an elliptic integral.
For the endpoint behaviour of distributions, we need only its behaviour near threshold, $M \rightarrow \mu_n \equiv \Sigma_i m_i$, which is given by \cite{BK}
\begin{multline}  \label{eqn:phs2}
\Phi_{n} (M \to P_1 \dots P_n) ~ \stackrel{M \to  \mu_n}{\longrightarrow} ~ \\ \frac{1}{2^{\frac{5n+7}{2}}\pi^{\frac{3n-5}{2}}\Gamma (\frac{3n-3}{2})} \left( \frac{\Pi_i m_i}{\mu_n^{3}}\right)^{\frac{1}{2}} (M - \mu_n)^{\frac{3n-5}{2}}.
\end{multline}

To compute the $m_T$ distribution for a single parent of mass $M$ decaying at rest to a single, massless, visible particle ($P$) and $n$, massless, invisible particles ($X$), we note that the definition of $m_T$ reduces, in this case, to
\begin{gather}
m_T^2 \equiv 4p_T^2.
\end{gather}
The differential phase space for the decay process $M \to P+nX$ can be conveniently written by taking the convolution of the two-body decay process with the system of $n$ invisible particles (collectively denoted as $I$), having invariant mass $m_I$
\begin{multline}
d\Phi_{n+1} (M \to P+nX) =  \\ \int_0^{M^2} \frac{dm_I^2}{2\pi} ~d\Phi_{2} (M \to P I)~
d\Phi_n (I \to nX).
\label{fact}
\end{multline}
Using (\ref{eqn:phs}), we obtain the normalized phase-space distribution in $m_T^2$, in the center-of-mass frame (or a frame boosted orthogonally to the transverse plane)
\beq \label{eqn:mtdist11}
\frac{M^2}{\Phi_{n+1}}\frac{d\Phi_{n+1}}{dm_T^2} (M\to P+nX) =A_n\left( \frac{m_T^2}{M^2}\right), 
\eeq
where 
\beq 
A_1(z)=\frac{1}{2\sqrt{1-z}}
\eeq
and
\beq \label{eqn:mtn14}
A_n(z) =\frac{n(n-1)}{2}\int_{0}^{1-\sqrt{z}} dx~ \frac{x^{n-2}}{\sqrt{(1-x)^2-z}},~~{\rm for}~n>1,
\eeq
such that
\beq
A_2(z) = \log \left( \frac{1+\sqrt{1-z}}{\sqrt{z}}\right),
\eeq
\beq
A_3(z) = 3\log \left( \frac{1+\sqrt{1-z}}{\sqrt{z}}\right) -3\sqrt{1-z},
\eeq
\beq
A_4(z) = 3\left( 2+ z \right)\log \left( \frac{1+\sqrt{1-z}}{\sqrt{z}}\right) -9\sqrt{1-z}, \; \&c.
\eeq

To compute the distribution with two visible particles, it is convenient to write the differential phase space as a convolution between the visible ($V=2P$) and invisible ($I=nX$) systems, having invariant masses $m_V$ and $m_I$, respectively,
\begin{multline}
d\Phi_{n+2} (M \to 2P+nX) =  \int_{m_V+m_I<M} \frac{dm_V^2dm_I^2}{(2\pi)^2} \\ d\Phi_{2} (M \to V I)~d\Phi_{2} (V \to 2P)~
d\Phi_{n} (I \to nX).
\end{multline}
The normalized phase-space distribution of the transverse mass  (\ref{mt}) of the two visible particles, in the center-of-mass frame (or boosted orthogonally to the transverse directions), is then
\beq \label{eqn:mtn241}
\frac{M^2}{\Phi_{n+2}}\frac{d\Phi_{n+2}}{dm_T^2}  (M\to 2P+nX) =B_n\left( \frac{m_T^2}{M^2}\right), 
\eeq
where 
\beq 
B_1(w) =\frac 12 \left[ 1- \frac{(1-2 w)}{\sqrt{w(1-w)}}\arctan \sqrt{\frac{w}{1-w}}\right] 
\eeq
and 
\begin{multline} \label{eqn:mtn24}
B_n(w) = n^2(n^2-1) \int_0^w dy\int_0^{(1-\sqrt{w})(1-y/\sqrt{w})}dx  \\ \frac{x^{n-2}(1-y^2w^{-2})}{2\sqrt{(1-x)^2-w-2xy+y^2(1-w^{-1})}},~~{\rm for}~n>1,
\end{multline}
such that
\bea
B_2(w) &=&2\sqrt{1-w} -(1-w)+4w \log \left( \frac{1-\sqrt{1-w}}{w\sqrt{w}}\right) 
\nonumber \\
&&-(1-4w)\sqrt {\frac{1-w}{w}} \arctan \sqrt{\frac{w}{1-w}}, \; \&c.
\eea

The normalized phase-space distribution in the invariant mass of the two visible particles is
\beq \label{eqn:minv2vni}
\frac{M^2}{\Phi^{(n+2)}}\frac{d\Phi^{(n+2)}}{dm_V^2}  (M\to 2P+nX)=C_n\left( \frac{m_V^2}{M^2}\right),
\eeq
where
\beq
C_1(a)= 2(1-a)
\eeq
and
\begin{multline}
C_n(a) = \\ n^2(n^2-1)\int_0^{(1-\sqrt{a})^2}dx~x^{n-2} \sqrt{(1+a-x)^2-4a}~~~{\rm for}~n>1,
\end{multline}
such that
\beq
C_2(a)= 6(1-a^2) +12a\log a,
\eeq
\beq
C_3(a)= 12(1-a)(1+10 a+a^2) +72a(1+a)\log a,
\eeq
\beq
C_4(a)= 20(1-a^2)(1+28 a+a^2) +240a(1+3a+a^2)\log a,  \; \&c.
\eeq

For a point-like decay of a single particle to two, massless, visible particles and an invisible particle of mass $m_X$, the $m_T$
distribution is given by
\begin{equation}\label{eqn:massive2+1}
\frac{M^2}{ \Phi_{3}}\frac{d \Phi_{3}}{d
  m_T^2}  (M\to 2P+X)=
D \left(\frac{m_T^2}{ M^2},\frac{m_X^2}{M^2}\right),
\end{equation}
where
\begin{widetext}
\begin{eqnarray}
\hspace{-0.1in}D\left(w,x\right)=\frac{1}{2\left(1-w\right)^2\left(1+2x\ln{x}-x^2\right)}
\begin{cases}
g\left(w,x\right)+\left(1-w-3x\right)\sqrt{\left(1-x\right)^2-2w\left(1+x\right)+w^2}\nonumber
\\\hspace{0.2in}-\frac{\left(1-x-w\right)\left(1-x-w\left(3+2x-2w\right)\right)}{\sqrt{w\left(1-w\right)}}\arccos{\left(-\sqrt{w}\frac{1+x-w}{1-x-w}\right)},
&  \textrm{for\,\,} \sqrt{w}\le 1-\sqrt{x},\\
g\left(w,x\right)-\frac{\left(1-x-w\right)\left(1-x-w\left(3+2x-2w\right)\right)}{\sqrt{w\left(1-w\right)}}\pi,
& \textrm{for \,\,} 1-\sqrt{x}<\sqrt{w}\le 1-x,
\end{cases}
\end{eqnarray}
and 
\begin{equation}
g\left(w,x\right)=3x\sqrt{\left(1-x\right)^2-w}+\frac{\left(1-x-w\right)\left(1-x-w\left(3+2x-2w\right)\right)}{\sqrt{w\left(1-w\right)}}\arccos{\left(-\frac{\sqrt{w}x}{1-x-w}\right)}.
\end{equation}
\end{widetext}
For a one-step cascade decay with two, massless, visible particles and a massless, invisible particle, the distribution of $m_T$ is given by
\begin{gather} \label{eqn:casc}
\frac{M^2}{ \Phi_2^2}\frac{d \Phi_2^2}{d m_T^2}  (M\to P+Y \to 2P+X) = E\left(\frac{m_T^2}{M^2}, \frac{m_Y^2}{M^2}\right),
\end{gather}
where
\begin{eqnarray} 
\hspace{-0.4in}E(y,z)  &\equiv& \frac{1}{2(1-z)\sqrt{y}\sqrt{1-y}} \times \nonumber\\
&&\int_a^{\sqrt{y}} dx \frac{x(2\sqrt{y}-x)}{(1+x\sqrt{y}-y)\sqrt{1-(\sqrt{y}-x)^2}}
\end{eqnarray}
and where
\begin{gather}
a=
\begin{cases}
0, &\text{if $y\leq 1-z$,} \\
\frac{y-\left(1-z\right)}{\sqrt{y}}, &\text{if $y > 1-z$.}
\end{cases}
\end{gather}

For events in which two parents, both of mass $M$, are produced at rest and each decays into a visible particle ($P$) and to $k$ or $l$ invisible particles, where $k+ l = n$, the normalized phase-space distribution in $m_{T2}^2$ is given by
\beq \label{eqn:mt2n14}
\frac{M^2}{ \Phi_{k+1} \Phi_{l+1} }\frac{d  (\Phi_{k+1} \Phi_{l+1}) }{d m_{T2}^2}  (2M\to 2P+(k+l)X)= F_{kl}\left( \frac{m_{T2}^2}{M^2}\right), %~~~~~{\rm for}~ 0\le m_{T2}^2 \le M^2,
\eeq
where 
\beq 
F_{kl} (z) =\frac{4}{\pi} \int_z^1 da~\int_{z/a}^1 db~\frac{ab }{\sqrt{z(ab-z)}}A_{k}(a^2) A_{l}(b^2),
\eeq
and the integrals $A_{n}(z)$ were defined above.
%%%%%%%%%%%%%%%%%%%%%%%%%%%%%%%%%%%%%%%%%%%%%%
\bibliography{countingdm}
%%%%%%%%%%%%%%%%%%%%%%%%%%%%%%%%%%%%%%%%%%%%%%
%%%%%%%%%%%%%%%%%%%%%%%%%%%%%%%%%%%%%%%%%%%%%%
\end{document}